\documentstyle[aps,epsfig]{revtex}

\begin{document}
\baselineskip=10pt
%--------------------------------------------------------------------------  

\def\mathrm{ }
% only for old LaTeX systems that don't recognize \mathrm!

\def\1/2{ {\scriptstyle{1\over2}} }

%--------------------------------------------------------------------------  
\def\Astrofisica{Astrof\'\i{}sica}
\def\Fisica{F\'\i{}sica}
%--------------------------------------------------------------------------
\title{Renormalization Group Analysis of a Quivering
String Model of Posture Control}
%--------------------------------------------------------------------------  
\author{ Francisco Alonso-S\'anchez$^{*}$ and David Hochberg$^{+,*}$
}
%--------------------------------------------------------------------------  
\bigskip
%--------------------------------------------------------------------------
\address{ 
$^{+}$Laboratorio de \Astrofisica\ Espacial y  \Fisica\
Fundamental, Apartado 50727, 28080 Madrid, Spain\\
$^{*}$Centro de Astrobiolog\'\i{}a 
(Associate Member of the NASA Astrobiology Institute), 
INTA, Ctra. Ajalvir, Km. 4, 28850 Torrej\'on de Ardoz, Madrid, Spain }
%--------------------------------------------------------------------------  
\bigskip
%--------------------------------------------------------------------------
%--------------------------------------------------------------------------  
\maketitle
%--------------------------------------------------------------------------  

\bigskip

{\small
  
  {\abstract} Scaling concepts and renormalization group (RG) methods
  are applied to a simple linear model of human posture control
  consisting of a trembling or quivering string subject to damping and
  restoring forces. The string is driven by uncorrelated white
  Gaussian noise intended to model the corrections of the
  physiological control system. We find that adding a weak quadratic
  nonlinearity to the posture control model opens up a rich and
  complicated phase space (representing the dynamics) with various
  non-trivial fixed points and basins of attraction.  The transition
  from diffusive to saturated regimes of the linear model is
  understood as a crossover phenomenon, and the robustness of the
  linear model with respect to weak non-linearities is confirmed.
  Correlations in posture fluctuations are obtained in both the time
  and space domain.  There is an attractive fixed point identified
  with falling.  The scaling of the correlations in the front-back
  displacement, which can be measured in the laboratory, is predicted
  for both the large-separation (along the string) and long-time
  regimes of posture control.

\bigskip


} 
% end-abstract

%------------------------------------------------------------------------------
\section{Introduction}
%------------------------------------------------------------------------------

A wide variety of systems subject to noise, random forces and
interactions can be studied in depth by means of non-equilibrium
statistical mechanics. This holds true whether the system in question
is fundamentally of a chemical, biological, or physical nature.  When
the phenomena under study admits mathematical modeling by means of
stochastic partial differential equations, many powerful techniques
can be used to analyze the effects that noise, fluctuations and random
disturbances have on the dynamics as one changes both the spatial and
temporal resolution scales at which the system is observed.  The
possibility to be able to use such techniques becomes especially
pressing given that many typical real systems of interest are
characterized by having many degrees of freedom interacting
nonlinearly, leading to the competition between different length and
time scales, with all scales evolving in the presence of noise and
subject to uncontrollable external effects and contingencies.  One of
these important techniques is provided by the renormalization group
(RG), suitably extended to dynamical systems and systems out of
equilibrium \cite{Ma,Cardy}. Some recent results of renormalization
group analyses of the kind presented in this paper have been obtained
for diverse phenomena ranging from stirred fluids \cite{FNS} and
turbulence \cite{Yakhot,Frisch}, to surface growth phenomena
\cite{KPZ,MHKZ,Sun-Plischke,Frey-Tauber,Barabasi}, flame front
propagation \cite{Cuerno-Lauritsen}, and cosmological large-scale
structure formation
\cite{Berera-Fang,Hochberg-Mercader,PGHL,BDGP,DHMPS}.

Fluctuations and noise are known to be present in physiological
systems as well.  Recently, a simple continuum model of human posture
control was proposed \cite{Chow-Collins} that captures the gross or
coarse-grained features underlying the physical mechanisms and adjusts
well to laboratory measurements of time-varying displacements of the
front-to-back (anteroposterior) sway recorded for human subjects in an
stationary upright stance \cite{Collins-DeLuca,Lauk-Chow}. Despite the
fact that the actual human postural control system must undoubtedly be
highly complex, the stochastic model introduced by Chow and Collins
\cite{Chow-Collins} is described by a linear, and hence, exactly
solvable, stochastic differential equation in one spatial dimension
(in the following, temporal derivatives are denoted by an overdot, and
the primes stand for spatial derivatives):
\begin{equation}\label{ChowCollins}
\beta \ddot y + \dot y - \nu y'' + \alpha y = \eta(x,t),
\end{equation}
where $y(x,t)$ denotes the time-dependent front-back displacement
measured with respect to the vertical located at $y=0$ (here we take
the $x$-axis to denote the vertical axis as $z$ will be reserved for
the dynamical exponent which we introduce and calculate below). This
is of course immediately recognized as a one-dimensional wave equation
subject to friction $(\dot y)$, a linear restoring or pinning force
$(\alpha y)$ and a stochastic or random force $(\eta)$. The
onset-of-damping time scale is set by $\beta$, the pinning time scale
by $\alpha^{-1}$ and $\nu$ is an effective string tension parameter.
(This continuum equation describes the motion of a quivering damped
string with a linear restoring force. A discrete version of this model
would involve a chain of coupled random walkers, or monomers, and in
this guise can be denoted as a pinned polymer \cite{Chow-Collins}).
The stochastic noise is taken to be Gaussian with zero mean. For
subsequent calculational purposes, it is convenient to characterize
the noise spectrum directly in Fourier space $(k,\omega)$:
\begin{eqnarray}\label{noise}
\langle \eta(k,\omega) \rangle &=& 0,\nonumber \\
\langle \eta(k,\omega)\eta(k',\omega') \rangle &=& \Gamma(k,\omega)(2\pi)^2
\delta(k+k') \delta(\omega + \omega'),
\end{eqnarray}
where the angular brackets denote averaging with respect to the noise,
and the noise spectrum function $\Gamma(k,\omega)$ may contain in
general both short range and long range correlations in space and/or
time. We consider white noise here, so the spectral function is
proportional to a constant noise amplitude: $\Gamma = 2{\cal A}$.
In \cite{Chow-Collins}, noise exhibiting short-time temporal
correlations but uncorrelated (or white) for long time scales was used
in computing the correlation and response functions directly from the
Fourier transform of the wave equation Eq.(\ref{ChowCollins}) where
the former were fit to laboratory posture data allowing a
phenomenological determination of $\beta,\alpha^{-1}$, an effective
noise amplitude and a parameter characterizing the short-time noise
correlations.

In \cite{Chow-Collins} it is claimed that nonlinearities are not
necessary to explain the posture data of healthy standing individuals,
though it was also recognized that this may not be the case for
subjects with balance disorders (nor even for normal individuals
subject to a sufficiently large perturbing ``kick''). These effects
clearly lie outside the scope of the linear model
Eq.(\ref{ChowCollins}) as do posture displacements outside the sway
envelope or cone of stability \cite{McCollum,Nasher}, and there is
clear motivation for extending that model to include weak nonlinear
terms. The minimal nonlinear term one can include in
Eq.(\ref{ChowCollins}) is of the form $\sim y^2$, which also serves to
break the $y \rightarrow -y$ symmetry in Eq.(\ref{ChowCollins}).
Including this quadratic term makes good sense from the physiological
point-of-view since real anteroposterior motion is intrinsically
asymmetric\cite{Chow-Collins}.  Physically, this amounts to having a
``pinning force'' that varies with the amplitude of the horizontal
displacement $y$. We will see, moreover, that such a term is needed in
order to account for falling.  In this paper we analyze the importance
and impact of weak nonlinearities and to check the robustness of the
linear model by means of a dynamical renormalization group analysis.
We therefore consider the following {\em nonlinear} stochastic wave
equation given by
\begin{equation}\label{nonlinear}
\beta \ddot y + \dot y - \nu y'' + \alpha y + \epsilon y^2 = \eta(x,t),
\end{equation}
where $\epsilon \geq 0$ is the strength of the quadratic nonlinearity
and we take the noise spectrum to be white (i.e., uncorrelated) at
large scales: that is $\Gamma(k,\omega) = 2{\cal A}$, where the noise
amplitude is denoted by $\cal A$, and serves also as a loop-counting
parameter \cite{HMPV-1}(an expansion of the solutions of
Eq.(\ref{nonlinear}) in loop diagrams is a convenient and powerful way
to organize the calculation.).
 
We are interested in the correlations in the solution(s) of
Eq.(\ref{nonlinear}) in the so-called hydrodynamic limit corresponding
to large spatial separations (along the vertical axis) and long time
intervals: $|x - x'| \rightarrow \infty$ and $|t - t'| \rightarrow
\infty$.  This will tell us how the posture fluctuations are
correlated along the length of the body at any given instant and how
they are correlated in time at any given point on the body.  In terms
of Fourier variables (momentum $k$ and frequency $\omega$) this limit
is taken by letting $(k,\omega) \rightarrow 0$. The scaling
information and universality class of this nonlinear wave equation is
contained in two critical exponents: the dynamic exponent $z$ and the
``roughness'' exponent $\chi$.  These exponents are first obtained via
a simple scale transformation of the stochastic equation of motion,
and forms part of the full RG transformation (a course-graining or
thinning-out of the degrees of freedom, followed by a re-scaling)
\cite{Ma}.  A change of space and time scales $x \rightarrow sx$, $t
\rightarrow s^z t$ is accompanied by a corresponding change of scale
in the displacement field variable $y \rightarrow s^{\chi} y$. Under
this scaling the stochastic equation of motion Eq.(\ref{nonlinear})
transforms according to
\begin{equation}
s^{-z} \beta  \ddot y + \dot y - s^{z-2} \nu  y'' + s^{z} \alpha  y +
s^{z+\chi} \epsilon  y^2 = s^{\frac{1}{2}(z-1) -\chi}\eta,
\end{equation}
where we have used the the noise two-point correlation function
Eq.(\ref{noise}) to determine the scaling of the noise source. Under
this transformation, the individual parameters appearing in
Eq.(\ref{nonlinear}) therefore scale as
\begin{eqnarray}\label{naive}
\beta &\rightarrow& s^{-z} \beta, \nonumber \\
\nu &\rightarrow& s^{z-2} \nu ,\nonumber \\
\alpha &\rightarrow& s^{z} \alpha,\nonumber \\
\epsilon &\rightarrow& s^{z+\chi} \epsilon,\nonumber \\
{\cal A} &\rightarrow& s^{z-2\chi-1} {\cal A}.
\end{eqnarray}
At a {\it fixed point} of this scaling, the model parameters and the
field $y$ no longer change under a re-scaling for certain specific
values of the exponents $z$ and $\chi$. The model parameters approach
their fixed-point values, $\beta \rightarrow \beta^*, \nu \rightarrow
\nu^*$, etc., and this fact gives rise to a corresponding fixed-point
equation of motion, which is Eq.(\ref{nonlinear}) written in terms of
the fixed point parameters. Thus, each fixed point corresponds to a
distinct dynamics governing the long time and large distance $(s
\rightarrow \infty)$ behavior of the model. The dynamical phase space
is thus divided or ``partioned'' into various domains or basins of
attraction (or repulsion), each domain associated with a given fixed
point. We can use this fixed point information to {\it predict} the
asymptotic scaling of the displacement correlation function in both
the temporal and spatial domains. This will be one of the main
objectives of this paper.

Independently of the RG, and in preparation for the results to be
obtained, it is useful to derive the general scaling form of the
correlation function of transverse displacements. Under a global scale
transformation, the displacement field transforms according to
\begin{equation}\label{scaling1}
y(sx,s^zt) = s^{\chi}\, y'(x,t),
\end{equation}
which merely states that under a space and time rescaling, the
displacement field can, and generally does, transform into a {\it
  distinct} (hence the prime) functional form $y'$, apart from picking
up an overall factor. Thus, a change of scale will generally change
the function itself, {\it unless} one is in the scaling or power law
regime. When the system is known to be in a scaling regime, then in
fact $y=y'$ and from Eq.(\ref{scaling1}), the auto-correlation
function therefore scales as
\begin{eqnarray}\label{corrscaling}
\langle y(x,t) y(0,0) \rangle &=& s^{-2\chi}\,\langle y(sx,s^zt) y(0,0) 
\rangle,\nonumber \\
&=& x^{2\chi}\, \Psi \Big( \frac{t}{x^z} \Big),
\end{eqnarray}
where (without loss of generality) we have chosen $s \sim x^{-1}$ and
$\Psi$ is a (dynamic) scaling function, which itself exhibits power
law behavior for asymptotic limits of its argument:
\begin{equation}\label{scaling2}
\Psi \big( u \big) = \{
\matrix{ A  & {\rm for} \qquad u \rightarrow 0,\cr
Bu^{\frac{2\chi}{z}} & {\rm for} \qquad u \rightarrow \infty,} 
\end{equation} 
for constants $A,B$.  The dynamic exponent $z$ describes the scaling
of relaxation times with length and $\chi$ is the ``roughness''
exponent of the string or polymer.  Thus, knowledge of these two
exponents is all that is required to determine the explicit scaling of
the correlation function within each dynamic phase of the model.  For
the linear model (i.e., $\epsilon = 0$) these exponents can be exactly
determined with little effort, and there are just two solutions (for
white noise).  For one, the wave equation for non-vanishing pinning
force $(\alpha \neq 0)$ is made scale-invariant with the choice $z =
0$ and $\chi = -\frac{1}{2}$. This exponent solution can be read off
directly from Eq.(\ref{naive}) taking into account the fact that the
noise amplitude is constant and hence, non-vanishing on all scales
$({\cal A} > 0)$. This immediately yields the exponent identity $z =
2\chi + 1$. For a finite fixed value of the pinning force, the only
possibility is to take $z=0$, since a positive $z > 0$ yields an
asymptotically divergent $\alpha$, while $z < 0$ would yield instead
an asymptotically vanishing value. We see that $\beta^* = \beta$ is
finite and the diffusion constant vanishes, $\nu^* = 0$. In this phase
then, there is no diffusion.  This exponent pair corresponds to the
experimentally observed scaling regime denoted as ``bounded'' or
``saturated'', and holds for the very latest times when the pinning
force has had time to correct for posture excursions from the vertical
and aligns the body in an upright stance \cite{Chow-Collins}. In the
earlier ``diffusive'' scaling regime, the pinning force has not had
sufficient time to act and is negligible, i.e., $\alpha \approx 0$,
and there is another exact exponent solution given by $z=2$ and $\chi
= \frac{1}{2}$, indicating that in this parameter regime, the model
belongs to the same universality class as the one-dimensional
Edwards-Wilkinson (EW) model \cite{Edwards}.  At this fixed point, we
have a finite diffusion $\nu^* = \nu$ and $\beta^* = 0$.  In this
phase, there is no wave propagation, since the second derivative in
time is absent.  Note that these simple scaling solutions have been
obtained from applying naive scaling arguments to the {\it linear}
equation.  However, as soon as the nonlinearity $\epsilon$ is turned
on, and no matter how weak, other nontrivial exponent solutions arise
for which the nonlinearity can become {\em relevant}.  The naive
scaling arguments are insufficient for obtaining the scaling exponents
in the fully nonlinear model.  The RG allows one to calculate $z$ and
$\chi$ in the combined presence of fluctuations and nonlinearities and
to calculate the exact asymptotic scaling of the correlation function
Eq.(\ref{corrscaling}) in all the basins of attraction.

The rest of this paper is organized as follows.  In Section II we make
use of a dynamic functional formalism for the perturbative calculation
of solutions to Eq.(\ref{nonlinear}) based on the Martin-Siggia-Rose
(MSR) Lagrangian. The bare correlation function, response function,
noise spectral function and bare interaction vertex function are
identified, their corresponding Feynman diagrams are introduced and
calculated and provide the basic elements of a systematic and
controlled loop expansion for the one-particle-irreducible (1PI)
diagrams which we then use for extracting the one-loop RG equations in
the low-energy regime.

In Section III we exhibit the set of nonlinear differential RG
equations for the dimensionfull parameters appearing in
Eq.(\ref{nonlinear}). For white noise these involve five equations:
one associated with each independent parameter appearing in the
equation of motion.  We then identify a convenient set of three
dimensionless couplings in terms of which these RG equations can be
expressed.  The RG flow is therefore represented in a
three-dimensional dimensionless parameter space and we solve for all
the one-loop fixed points in terms of this reduced set.  In this way
we find a total of two trivial fixed lines and four non-trivial fixed
points.  Linearization of the RG about each fixed point (or line)
reveals the {\it nature} of the fixed point, in the dynamical systems
sense (whether the fixed point is a source, a sink, a limit cycle, a
spiral, a saddle point, etc.) and yields linear stability information
which we quote in terms of the eigenvalues and eigenvectors of the
linearized RG.  The two lines of fixed points correspond to the
diffusive and bounded phases of the strictly linear model, and are
present in the non-linear model for all values of $\alpha$ and
$\beta$.  Of the four non-trivial fixed points, one is a stable spiral
which represents the ``falling'' phase. The other three are saddle
points which seem to have rather little influence on the long-range
and long-time dynamics, however.

Substitution of the fixed points back into the original set of RGs
yields the values of the critical exponents $z,\chi$ for each fixed
point and hence determines the exact asymptotic scaling properties of
the correlations Eq.(\ref{corrscaling}) in the basin of attraction (or
repulsion) of each fixed point, which is presented in Section IV.  The
numerical analysis of the fixed points is then repeated using an
alternative set of three dimensionless parameters suitable for
investigating the small $\alpha$ limit (corresponding to the diffusive
regime).  The use of this second set of parameters in conjunction with
the first is necessary in order to completely {\it cover} the entire
model parameter space.

The detailed structure of the nonlinear RG flow is revealed by
plotting the fixed points and mapping the numerically computed (and
normalized) vector field of the non-linear RG flow in the
neighborhoods of all the points in Section V.  Many important aspects
of the morphology of the dynamic phase space are qualitatively
revealed and allow conclusions to be drawn regarding the impact of the
weak nonlinearity.  This provides revealing information regarding the
shape and structure of the basins of attraction and complements the
analytic analysis.  Summary and conclusions are drawn in Section VI.

A number of explicit analytic calculational details needed for the
derivation of the RG equations are relegated to the Appendices. The
complete calculations leading to the one-loop response function are
presented in Appendix A, and similar calculations for the noise
spectral function and vertex renormalizations are given in Appendix B
and C, respectively.

%----------------------------------------------------------------------------
\section{Dynamic Functional Formalism}
%----------------------------------------------------------------------------

In this Section we make use of a functional integral representation of
non-equilibrium stochastic dynamics. This leads to the efficient
identification and extraction of the calculational elements (and
Feynman rules) needed for the perturbative calculation of the
solutions of any stochastic partial differential equation.  It is well
known how to map stochastic ordinary or partial differential equations
with additive noise into equivalent generating functionals
\cite{Rivers,Zinn-Justin}.  Essentially, there are two formally
distinct but physically equivalent routes one may follow, an option
one has at least in the case of Gaussian noise.  In the
Martin-Siggia-Rose (MSR) formalism
\cite{MSR,De-Dominicis-Peliti,Janssen} one introduces a fictitious
conjugate field (call it $\hat y$) with its own source term. The
equation of motion, in this case given by Eq.(\ref{nonlinear}), is
imposed as a constraint on the dynamic functional, and is realized
linearly.  In the minimal formalism, no conjugate field is introduced,
leading to a nonlinear realization of the constraint
\cite{Zinn-Justin,HMPV-1}.  For Gaussian noise in the minimal
formalism, the constraint appears quadratically in the argument of the
functional, while for non-Gaussian noise, it is inherits whatever
nonlinearities are present in the noise probability distribution
function itself \cite{HMPV-1}.  Here we develop the calculation
following the MSR approach, since this leads to a simpler structure
for the associated Feynman diagrams, and our immediate aim is to
obtain a perturbation expansion which can be set up, organized and
calculated in terms of a few elementary diagrams or graphs.

The MSR dynamic generating functional corresponding to
Eq.({\ref{nonlinear}) is given by (taking a translationally invariant
  noise spectrum)
\begin{equation}\label{functional}
Z[J,\hat J] = \int [dy][d\hat y] \, 
\exp \Big( -\1/2\int dx dt \,\, \hat y \Gamma \hat y
+i\int dx dt \,\, \hat y \left\{ \beta \ddot y + \dot y - \nu y'' + \alpha y
+ \epsilon y^2 \right\} + \int dx dt \, (yJ + \hat y \hat J) \Big),
\end{equation}
where $\hat y$ denotes the conjugate field and $J,\hat J$ are
arbitrary sources for $y$ and $\hat y$, respectively.  The noise
$\eta$ has been integrated out exactly, and appears in this functional
only through its two-point or correlation function $\Gamma$.  There is
also in principle a certain Jacobian determinant factor in passing
from Eq.(\ref{nonlinear}) to Eq.(\ref{functional}), but it can be
shown on general grounds to be a constant, and hence irrelevant for
computing normalized correlation functions (see, e.g.,
\cite{Zinn-Justin,HMPV-1}).  The noise spectrum $\Gamma$ as written
here is understood to be given in terms of $x,t$.  All the dynamic and
fluctuation information contained in Eq.(\ref{nonlinear}) is also
contained in $Z$, which is an alternative representation of the
dynamics.  In preparation for the RG transformation, which is most
straightforwardly implemented in the Fourier domain, we cast this
functional in terms of momentum and frequency variables from the
outset.

To this end, we introduce Fourier transforms for the physical and
conjugate fields and the noise spectrum, i.e.,
\begin{equation}
y(x,t) = \int^{>}\frac{dk}{2\pi}
\int_{-\infty}^{\infty}\frac{d\omega}{2\pi}\,\, y(k,\omega) \, e^{i(kx-\omega t)},
\end{equation}
where, in a mild abuse of notation, we distinguish the functions from
their Fourier transforms only through their arguments; this however
avoids a clutter of notation later on.  Note we implicitly cut-off the
momentum integration at the scale $\Lambda = \frac{2\pi}{a}$ where $a$
plays the role of a minimum distance of spatial resolution or lattice
spacing. The cut-off symbol on the integral $(>)$ means one is to
integrate over all momenta in a ``shell'' such that $\Lambda/s < |k| <
\Lambda$ where $s > 1$ (see Eq. {\ref{domain}}).  The cut-off defines
the spatial scale above which it makes sense to use continuum
equations for modeling.  The quadratic or Gaussian part of the
functional (i.e, $\epsilon = 0$) can be exactly computed and serves as
the starting point for a perturbative expansion which we will use to
calculate the RGE's associated with the nonlinear stochastic wave
equation Eq.(\ref{nonlinear}).  {F}rom standard Gaussian integrations
\cite{Rivers,Zinn-Justin}, we have that (up to an overall irrelevant
constant prefactor)
\begin{equation}
Z_0[J,\hat J] = \exp \Big(
\int^{>}\frac{dk}{2\pi}
\int_{-\infty}^{\infty}\frac{d\omega}{2\pi}\, \left\{ \frac{\1/2 J(k,\omega)
\Gamma(-k,-\omega) J(-k,-\omega)}{\omega^2 + [\nu k^2 -\beta \omega^2
+\alpha]^2} + i\frac{\hat J(k,\omega) J(-k,-\omega)}{
i\omega -\beta \omega^2 + \nu k^2 + \alpha} \right\} \Big).
\end{equation}
{F}rom Eq.(\ref{functional}), it is clear that all noise-averages of
arbitrary products of physical fields and conjugate fields at distinct
points and times are obtained from the appropriate functional
derivatives of $\ln(Z)$ with respect to the source terms $J,\hat J$,
taking the sources to zero at the end of the calculation. We may thus
obtain the ``bare'' or zeroth order auto-correlation and response
functions directly in Fourier space (the zero ($0$) subscript denotes
the zero-coupling limit $\epsilon = 0$) as follows:
\begin{eqnarray}
\langle y(p_1,\omega_1)y(p_2,\omega_2) \rangle_0 &=& 
\frac{(2\pi)^4}{Z_0[J,\hat J]} \frac{\delta^2 Z_0[J,\hat J]}{\delta J(p_1,\omega_1)
\delta J(p_2,\omega_2)}|_{J=\hat J = 0}\nonumber \\
&=& (2\pi)^2\delta(p_1+p_2)\delta(\omega_1+\omega_2)
\frac{\Gamma(p_1,\omega_1)}{\omega_1^2 + [\nu p_1^2 -\beta \omega_1^2 
+\alpha]^2},\nonumber \\
&\equiv& (2\pi)^2\delta(p_1+p_2)\delta(\omega_1+\omega_2)C(p_1,\omega_1);
\end{eqnarray}
for the auto-correlation function and

\begin{eqnarray}\label{response}
\langle y(p_1,\omega_1)\hat y(p_2,\omega_2) \rangle_0 &=& 
\frac{(2\pi)^4}{Z_0[J,\hat J]} 
\frac{\delta^2 Z_0[J,\hat J]}{\delta J(p_1,\omega_1)
\delta \hat J(p_2,\omega_2)}|_{J=\hat J = 0}\nonumber \\
&=& i \frac{(2\pi)^2\delta(p_1+p_2)\delta(\omega_1+\omega_2)}{
i\omega_1 -\beta \omega_1^2 + \nu p_1^2 + \alpha},\nonumber \\
&\equiv& (2\pi)^2\delta(p_1+p_2)\delta(\omega_1+\omega_2)i \Delta(p_1,\omega_1),
\end{eqnarray}
for the response function.  The final equations in each case define
the corresponding {\it reduced} auto-correlation and response
functions: $C$ and $\Delta$, respectively. The delta function factors
reflect the fact that momentum and energy (or frequency) are conserved
as they ``flow'' through the correlation and response functions.  It
will be noted that the reduced response and correlation functions are
related via $C = \Delta^* \Gamma \Delta = \Gamma |\Delta|^2$, and this
relation will be used to obtain the noise amplitude renormalization
from knowledge of the correlation function. They also automatically
satisfy the fluctuation-dissipation theorem: $C(p,\omega) =
\frac{\Gamma}{\omega}\, {\rm Im} \Delta(p,-\omega)$.  The remaining
element from which the complete perturbative expansion of the dynamic
functional is developed is provided by the vertex function, which
necessarily involves the trilinear interaction term.  The complete
dynamic functional Eq.(\ref{functional}) can be written as
\begin{equation}\label{fullfunctional}
Z[J,\hat J] = \exp \Big(S_I[\frac{\delta}{\delta J},\frac{\delta}{\delta \hat J}]
\Big) \, Z_0[J,\hat J] = \sum_{n=0}^{\infty} \frac{(S_I)^n}{n!} \,
Z_0[J,\hat J],
\end{equation}
where the exponential of the following interaction operator acts on
the Gaussian part $(Z_0)$ of the generating functional:
\begin{equation}\label{barevertex}
S_I[\frac{\delta}{\delta J},\frac{\delta}{\delta \hat J}] =
i\epsilon \int \left\{ \prod_{j=1}^3 \frac{dk_j\,d\omega_j}{(2\pi)^2}\right\}  \,
(2\pi)^2\delta(k_1+k_2+k_2)\delta(\omega_1+\omega_2+\omega_3)\,
(2\pi)^6\frac{\delta^3}{\delta \hat J(k_1,\omega_1)\delta J(k_2,\omega_2)
\delta J(k_3,\omega_3)},
\end{equation}
and merely reflects the fact that the non-linear interaction term for
the MSR functional is a {\it cubic}-interaction involving one
conjugate field and two factors of the physical field: $\sim i\epsilon
\hat y y^2$.  The delta functions are a consequence of the energy and
momentum conservation at the vertex, and the strength of the vertex is
given by the parameter $\epsilon$, which in fact defines the reduced
vertex.  We complete the specification of the elements encountered in
a perturbative expansion of $Z$ by specifying the vertex function,
which we read off simply by inspection of Eq.(\ref{barevertex}):
\begin{equation}\label{epsilon}
\epsilon(k_1,k_2,k_3;\omega_1,\omega_2,\omega_3) = 
(2\pi)^2\delta(k_1+k_2+k_2)\delta(\omega_1+\omega_2+\omega_3)i \epsilon.
\end{equation}

Each of these calculational elements, $C,\Delta,\Gamma,\epsilon$, can
be represented by a simple diagram as depicted in Figure
(\ref{basicFrules}), and then used to systematically construct a
perturbation expansion of the dynamic functional, and its Legendre
transform (or effective action) which is the generator of the
one-particle irreducible (1PI) diagrams.  (Note: we remove the factor
of $i$ appearing in the bare vertex (\ref{barevertex}) and reponse
functions (\ref{response}) by re-defining the conjugate field $\hat y
\rightarrow i \hat y$ in the functional integral.)  The Feynman diagram
for the bare response function $\Delta$ is shown in Figure 1A, which
reflects the fact that the response function is constructed from the
mixed product of the conjugate field (wiggly line segment) times the
physical field (straight line segment). The arrow indicates the
direction that momentum $k$ and frequency $\omega$ flow through this
diagram (from the conjugate to the physical field) and is the
convention we adopt here. It is important to keep this in mind since
$\Delta(k,\omega) \neq \Delta(-k,-\omega)$, as can be seen by
inspection of (\ref{response}).  The bare correlation function is
shown in Figure 1B. As this is an even function in both momentum and
frequency, we need not indicate a flow direction for this function.
Finally, the bare trilinear vertex is shown in Figure 1C. This
involves one conjugate field and two physical fields. The bare
interaction is represented by a small open circle. The arrows indicate
the flow direction for the two physical fields and is just a
convention; the crucial point is that momentum and frequency are
conserved at each vertex, as per (\ref{epsilon}). These are used to
build up the 1PI diagrams, which are those diagrams that cannot be
broken down into disconnected sub-diagrams by cutting an internal line
(see examples below).  The class of 1PI diagrams are of fundamental
importance in any perturbative scheme based on diagrams since all
other diagrams can be unambiguously built up from these primitive
``building blocks''. In Figures (\ref{response1}),
(\ref{vertex1}),(\ref{noise1}), we depict the 1PI one-loop diagrams
contributing to the noise spectrum, response function and vertex
function. These diagrams are built up from the elementary vertex and
response and correlation functions. Note that the bare correlation
function is itself a {\it composite} function built up from the noise
spectrum and response functions and is {\it not} 1PI. This fact is
also reflected mathematically in the fact that the bare correlation,
noise and response functions obey a fluctuation-dissipation theorem.
The complete mathematical transcription of the diagrams is carried out
in the Appendices.
 
Finally, we note that each loop diagram is multiplied by an overall
symmetry factor. This factor receives a contribution from the
factorial coming from expanding out the exponential prefactor in
(\ref{fullfunctional}) to a given order in the coupling. For example,
a graph with $n$ bare interaction vertices yields a factor of
$\frac{1}{n!}$. The other contribution comes from simply counting the
number of distinct ways in which the given diagram can be built out of
elementary Feynman graphs (Fig. 1A-C) keeping the topological
structure fixed \cite{Amit}. Multiplying these two numbers together
yields the final net symmetry factor. For the response, noise and
vertex one-loop diagrams, these turn out to be $S_R = 4, S_N = 2$ and
$S_V = 4$, respectively.

\newpage
\begin{figure}[b]
\epsfysize=6cm 
{\centerline{\epsfbox{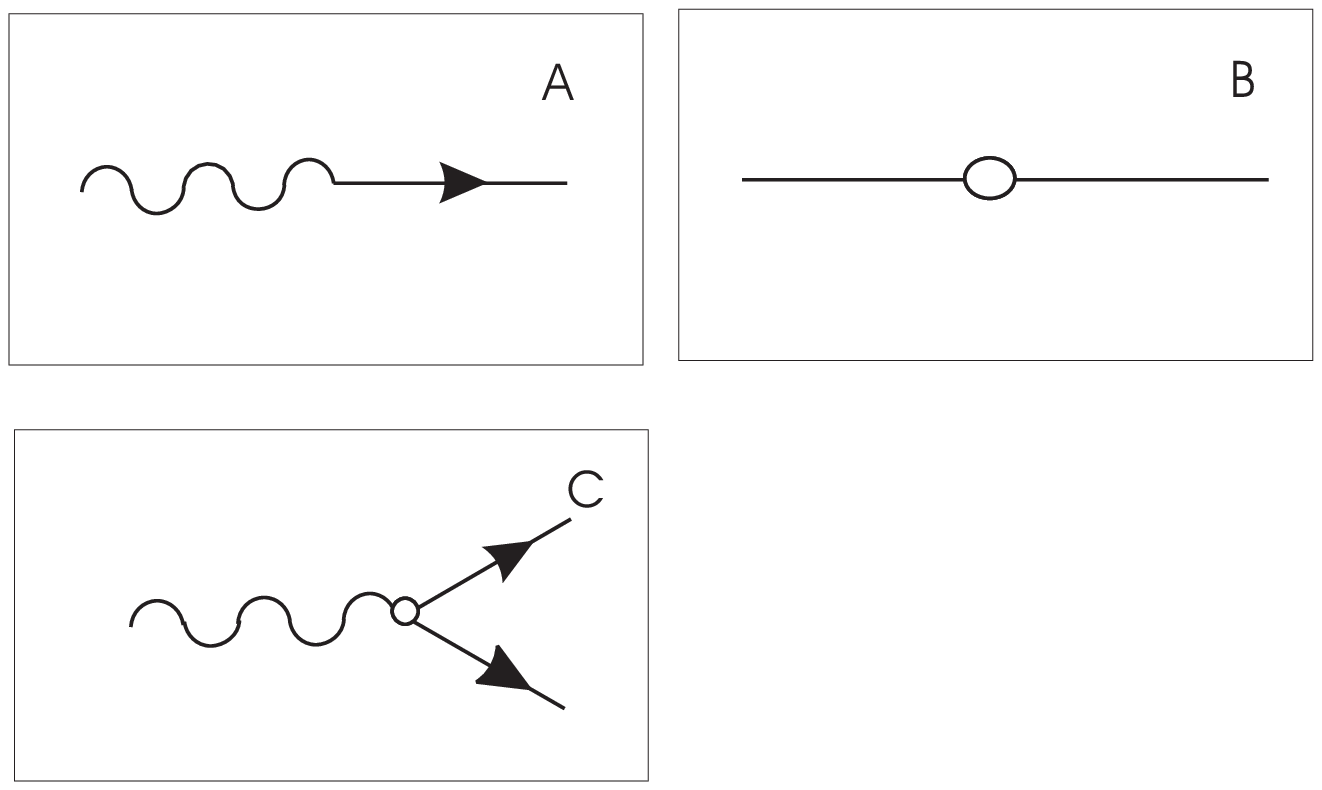}}}
\caption[]{\label{basicFrules}
A: The bare response function, B: The bare correlation function, C: 
The bare vertex function.}
\end{figure}

\begin{figure}[b]
\epsfysize=1.75cm
{\centerline{\epsfbox{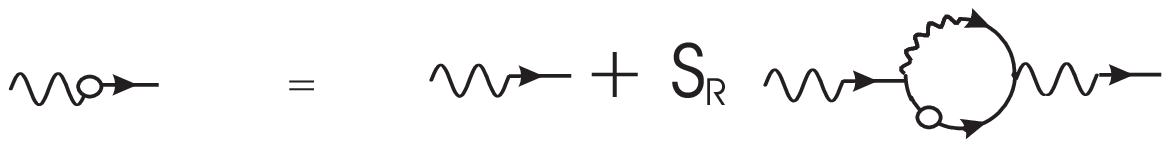}}}

\caption[]{\label{response1}
The one-loop corrected response function.}
\end{figure}

\begin{figure}[b]
\epsfysize=5cm 
{\centerline{\epsfbox{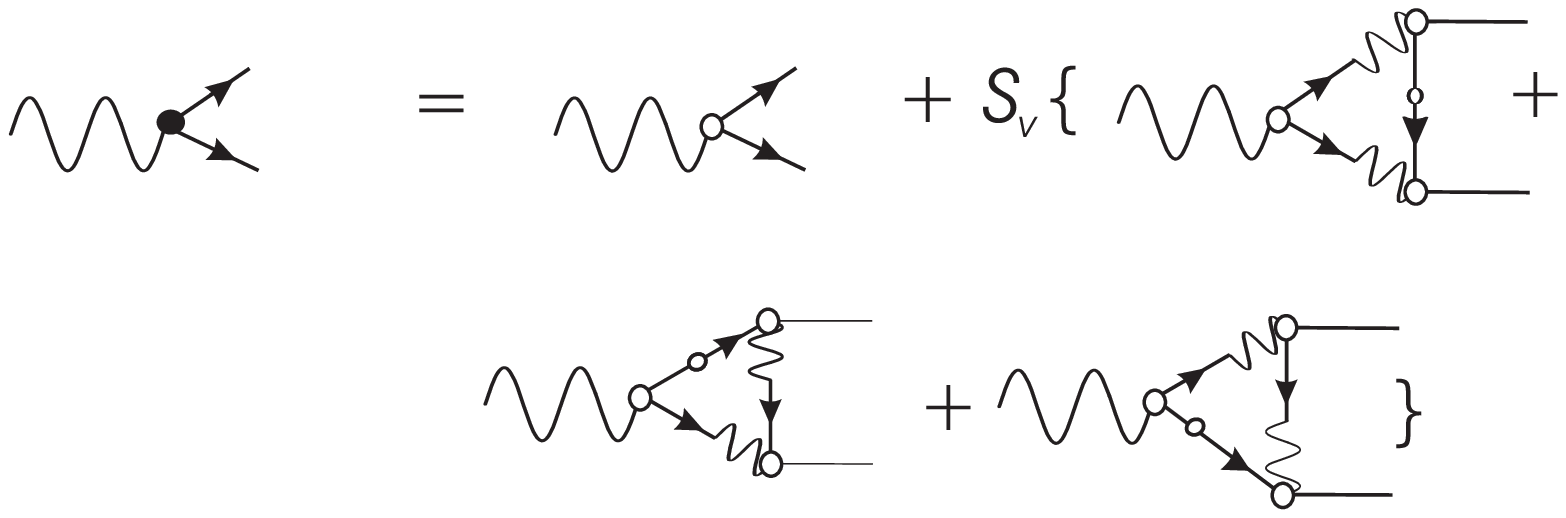}}}
\caption[]{\label{vertex1}
The one-loop corrected vertex function.}
\end{figure}

\begin{figure}[b]
\epsfysize=1cm 
{\centerline{\epsfbox{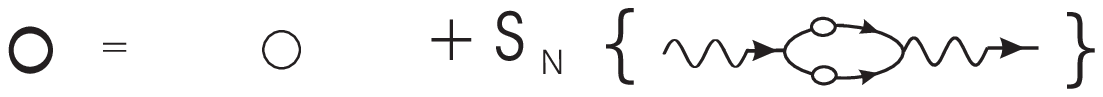}}}
\caption[]{\label{noise1}
The one-loop corrected noise spectral function.}
\end{figure}

\newpage

%----------------------------------------------------------------------------
\section{Dynamic Renormalization Group Equations and their Fixed Points}
%----------------------------------------------------------------------------

By means of the diagrammatic expansion, we calculate the one-loop
corrections to the response function $\Delta$, the noise spectral
function $\Gamma$ and the nonlinear vertex $\epsilon$ and from them,
obtain the associated one-loop corrections to the model parameters
$\alpha,\beta,\nu,{\cal A}$ and $\epsilon$ in the large-distance and
long-time limit.  The complete one-loop expressions are derived and
calculated explicitly in the Appendices A, B and C for the response
function, noise spectral function and the interaction vertex,
respectively. With these in hand, the next task is to carry out the
renormalization group transformation, leading to the differential flow
equations, whose steps we briefly review here \cite{Ma}.

\begin{enumerate}
\item{We first perform an infinitesimal Kadanoff or
    ``block''-transformation: that is, we integrate over a thin
    momentum shell $\Lambda/e^{\ell} \leq |k| \leq \Lambda$ where
    $\ell = 1 + \delta$, $0 < \delta << 1$. This means we integrate
    over (small-scale) fluctuations characterized by having their
    momentum in this range: $(\int_{-\Lambda}^{-\Lambda/e^{\ell}} +
    \int_{\Lambda/e^{\ell}}^{\Lambda}) \frac{dk}{2\pi}.$ Physically,
    this step serves to thin out the degrees of freedom
    (coarse-graining) and reduces the spatial resolution of the
    system. Note that in one space dimension, the ``momentum shell''
    reduces to two disjoint intervals. For $d \geq 2$ and higher, it
    is a thin spherical shell in momentum space.}
\item{ After performing this step, the resulting equations have a new,
    lower momentum cutoff of $\Lambda/e^{\ell}$. This means we have
    changed the lattice constant of the system, so to restore it to
    its original size, we re-scale the momenta by putting $k
    \rightarrow ke^{-\ell}$. This is identical to the scaling carried
    out earlier with $s = e^{\ell}$. The parameters are rescaled (but
    here, differentially) as in (\ref{naive}), but with additional
    corrections coming from the momentum-shell integrations carried
    out in step (1).}
\end{enumerate}
Applying these steps to the one-loop expressions (\ref{responseprime},
\ref{noiseprime}, \ref{vertexprime}), we obtain the following set of
differential renormalization group flow equations:
\begin{eqnarray}\label{RGEs}
\frac{d\alpha}{d\ell} &=& \alpha \Big(z - \frac{2 g}{(1+h)^2} \Big),\nonumber \\
\frac{d\beta}{d\ell} &=& \beta \Big(-z - \frac{g}{(1+h)^2}
\left\{ f+ \frac{1}{2f(1+h)^2} \right\}  \Big),
\nonumber \\
\frac{d{\cal A}}{d\ell} &=& {\cal A} \Big(z-2\chi -1 + \frac{4g}{4f(1+h)-1}
\left\{\frac{f^2}{(1+h)} + \frac{3f}{4(1+h)^2}-\frac{1}{4(1+h)^3}\right\} \Big),
\nonumber \\
\frac{d\epsilon}{d\ell} &=& \epsilon \Big(z + \chi + \frac{2g}{(1+h)^2}
\left\{ \frac{1}{(1+h)} -f + \frac{1}{4f(1+h) - 1}[
4f^2(1+h) + 3f - \frac{1}{(1+h)}] \right\} \Big),\nonumber \\
\frac{d\nu}{d\ell} &=& \nu \Big(z-2 +\frac{2g}{4f(1+h)-1}\left\{
[2 -\frac{8hf}{4f(1+h) - 1}]\Big(-\frac{f^2}{(1+h)} + 
\frac{5f}{4(1+h)^2} -\frac{1}{4(1+h)^3} \Big) \right.
\nonumber \\
&+& \left. 
4h \Big( \frac{3f^2}{2(1+h)^2}-\frac{f}{(1+h)^3} + 
\frac{1}{4(1+h)^4} \Big) \right\} \Big),
\end{eqnarray}
where the three {\it dimensionless} parameters $g,h,f$ are defined by
\begin{equation}\label{dimensionless}
g = \frac{\epsilon^2 {\cal A} \Lambda}{\pi \alpha^3},\qquad
h = \frac{\nu \Lambda^2}{\alpha}, \qquad
f= \alpha \beta.
\end{equation}
This reduction from five dimensionfull to three dimensionless
parameters is a concrete realization of a more general result from
dimensional analysis known as Buckingham's $\Pi$-Theorem
\cite{Buckingham}, which states that if one has $m$ dimensionfull
variables in a theory involving $n$ fundamental units (such as length,
time, mass, etc.)  then there exist $m - n$ independent dimensionless
groups or combinations of the $m$ original quantities. In our case, $m
= 5$, and $n = 2$, since all the parameters appearing in the
stochastic equation (\ref{nonlinear}) can be expressed in terms of two
fundamental units: namely length and time.  Thus, we expect to be able
to write the five RG equations exclusively in terms of $5-2 = 3$
independent dimensionless groups or combinations of the original
dimensionfull parameters.  The fact that a cut-off is introduced from
the RG transformation presents no problem since $[\Lambda] = L^{-1}$
has units of inverse length. The number of fundamental units remains
the same.  Below we introduce a second, independent group of three
dimensionless parameters which when taken with this first set, will
suffice to {\it cover} the entire parameter space.

The dimensionless coupling controlling the nonlinearity is $g$, and a
non-zero value of this coupling is what drives the RG flow away from
the free, or Gaussian limit. Since $g \propto \epsilon^2 {\cal A}$, to
have nontrivial flow simultaneously requires both a non-vanishing
nonlinearity $\epsilon \neq 0$ {\it and} fluctuations ${\cal A} > 0$.
Indeed, by setting $g = 0$ in (\ref{RGEs}), we immediately recover the
naive scaling laws given by (\ref{naive}), expressed here in
differential form, with $s = e^l$. Thus we see that the model
parameters depend on scale in a complicated way when nonlinearities
$(g \neq 0)$ are present.  In terms of these dimensionless variables,
the RG flow is given by the following set of three differential
equations:
\begin{eqnarray}\label{dimensionless1}
\frac{dg}{d\ell} &=& g \Big(-1 + \frac{g}{(1+h)^3}[15 + f(1+h) +6h] \Big), 
\nonumber \\
\frac{dh}{d\ell} &=& h \Big(-2 + \frac{g}{(1+h)^4}[3+3h+2h^2+f(-1+2h+3h^2)]\Big),
\nonumber \\
\frac{df}{d\ell} &=& -\frac{g}{2(1+h)^4} 
\Big(1 + 4f(1+h)^2 + 2f^2 (1+h)^2 \Big).
\end{eqnarray}
These are obtained by differentiating (\ref{dimensionless}) and using
(\ref{RGEs}).  Note the dependence on the two exponents has dropped
out.  We solve for the complete set of real fixed points whose
coordinates are $(g^*,h^*,f^*)$.  These are labelled, collected and
presented in Table I along with their associated critical exponents,
infrared eigenvalues, eigenvectors and the nature of the fixed point.
There is actually an entire line of attractive fixed points, denoted
by L1, located at $(0,0,f)$ , where $f$ is any real number.  A fit of
the linear model to the data yields $f^* = 0.05$. As we will see, $L1$
is associated with the final saturated phase for which $\nu^* = 0$.
This implies that $h^* = 0$. In the linear model $g = g^* = 0$, so the
posture data picks out a distinguished point on the line $L1$, and we
see the final phase of posture control is controlled by an infrared
attractive fixed point at $(0,0,f^*)$.  The quadratic term gives rise
to three saddle points, P3, P4, and P5, and one stable spiral, P2;
their coordinates are listed in the second column of Table I.  When we
substitute these fixed points into the original set of RGE's
(\ref{RGEs}), we obtain the corresponding pair of critical exponents:
$z=z(g^*,h^*,f^*),\chi= \chi(g^*,h^*,f^*)$ associated with each fixed
point (or fixed line):
\begin{equation}\label{exponents}
z = \frac{2g^*}{(1+h^*)^2} \qquad 
\chi = -\frac{(1-g^*[3 + f^*(1+h^*) + 2h^*] + h^*[3+3h^*+{h^*}^2] )}
{2(1+h^*)^3},
\end{equation}
which we obtain from using the pair of equations for ${\cal A}$ and
$\alpha$ in (\ref{RGEs}) to solve for $z,\chi$. Any pair of equations
depending on $z$ and $\chi$ taken from (\ref{RGEs}) can be used to
solve for these exponents provided that neither equation in the pair
{\it trivially} evaluates to zero at the fixed point whose exponents
one wishes to compute (thus, for example, if $\beta^* = 0$ , then we
cannot use the equation $d\beta/dl = 0$ to solve for $z$, and so
forth.).  Taking this obvious restriction into account, we have
checked that the exponents calculated taking {\it all} possible pairs
of the RGE's agree. This serves as an important consistency check of
the entire calculation. The exponents are listed in the third column
of Table I.  By linearizing the RG equations (\ref{dimensionless1})
around each fixed point, we obtain additional information regarding
the stability of the fixed point and can classify the infrared
stability properties of the point. This information is helpful for
visualizing the RG flow and interpreting the flow graphs in Section V.
At each fixed point we substitute $g = g^* + \delta g, h = h^* +
\delta h, f = f^* + \delta f$, into (\ref{dimensionless1}) retain all
terms up to order $O(\delta g,\delta h,\delta f)$ and compute the
eigenvectors and eigenvalues of this linear system.  This information
completely characterizes the RG flow in the linearized neighborhood of
each fixed point. The eigenvalues and eigenvectors are listed in
columns four and five of Table I.  The unit direction vectors in this
coupling space have components $\hat g = (1,0,0), \hat h = (0,1,0),
\hat f = (0,0,1)$. Note that the line $L1$ has only two RG flow
eigenvectors associated with it.  {F}rom knowledge of the eigenvalues,
we immediately predict the nature of the fixed point, and this is
given in the last column of Table I.

\widetext 
\begin{table}
\begin{center}
\begin{tabular}{|c|c|c|c|l|l|} 
\hline
{\rm Fixed}&{\rm Position}        &Critical exponents&{\rm IR-eigenvalues}        &{\rm IR-eigenvectors} &{\rm Class}     \cr
{\rm Point}&$ (g*,h^*,f^*)$       &$(z,\chi)$        &                            &                 &                     \cr
\hline \hline 
L1         & $(0,0,f)$            & $ (0,-1/2)$      & $(-2,-1,0)$                &$v^{1}_{L1}=(1,0,0)$&{\rm Stable line} \cr
           &                      &                  &                            &$v^{2}_{L1}=(0,1,0)$&{\rm attractor}   \cr 
\hline
P2         & $(-0.37,-2.15,-1.79)$& $(-0.56, -0.41)$ &$(-6.79, -0.67 \pm 0.27 i) $&$v^{1}_{P2}= (-0.17,-0.99,0.03)$&{\rm Stable spiral}\cr
           &                      &                  &                            &$v^{2}_{P2}= (0.57+0.11i, 0.72,-0.24- 0.28i)$&\cr
           &                      &                  &                            &$v^{3}_{P2}= (0.57-0.11i, 0.72,-0.24+ 0.28i)$&\cr
           &                      &                  &                            &                 &                     \cr
P3         & $(0.07, 0, -0.29)$   & $(0.14, -0.41)$  & $(-1.78,1,-0.10 )$         &$v^{1}_{P3}= (0.06,1.00,-0.04)$&{\rm Saddle point}  \cr
           &                      &                  &                            &$v^{2}_{P3}= (1.00,0.00,0.00)$&                \cr
           &                      &                  &                            &$ v^{3}_{P3}= (0.00,0.00,1.00)$&                 \cr
           &                      &                  &                            &                 &                     \cr
P4         & $(0.08, 0, -1.71)$   & $(0.15, -0.45)$  & $(-1.65, 1.,0.11)$         &$v^{1}_{P4}= (0.08,1.00,-0.04)$&{\rm Saddle point}   \cr
           &                      &                  &                            &$v^{2}_{P4}= (1.00,0.00,0.00)$&                     \cr
           &                      &                  &                            &$v^{3}_{P4}= (-0.01,0.00,1.00)$&            \cr
           &                      &                  &                            &                 &                     \cr
P5         &$(2.86, -2.96, -0.07$)& $(1.48, 0.03)$   & $(-8.24, 1.50, -1.28)$     &$v^{1}_{P5} = (0.21,0.98,0.01)$&{\rm Saddle point}\cr  
           &                      &                  &                            &$v^{2}_{P5}= (0.97,-0.23,0.01)$&       \cr
           &                      &                  &                            &$v^{3}_{P5}= (0.79,0.45,-0.41)$   &         \cr
\hline
\end{tabular}
\caption{}{\label{tab:one}Fixed Points in terms of ${g,f,h}$.}
\end{center}
\end{table}
Of the four non-trivial fixed points, three of them $(P2,P3,P4)$ are
well within the limits of perturbation theory as they all correspond
to small values of the dimensionless coupling $|g^*| < 1$.  By
contrast, the coupling for the $P5$ fixed point turns out to be rather
large: $g^* = 2.86 > 1$, and it is likely this point is an artifact of
perturbation theory. In any event, as can be seen from inspection of
the graphs in Section V, $P5$ is ``far'' away from the remaining fixed
points and has very little or no impact on the RG flow in the small
coupling region, which is the region safely explored by perturbation
theory. Both $P3$ and $P4$ are near the point $(0,0,f^*)$ belonging to
$L1$, but a study of the flow in the neighborhood of $(0,0,f^*)$ shows
these to be unimportant and substantiates the claim made in
\cite{Chow-Collins} that nonlinear effects are not needed to explain
posture data of {\it healthy} individuals. Nevertheless,
nonlinearities are needed to account for falling, and the stable
spiral $P2$ is an attractor for ``falling'', as we now show.  For $P2$
the fixed coupling $g^* = -0.37$ is {\it negative}.  But $g^* < 0
\Longrightarrow \alpha^* < 0$, since $\epsilon^2 > 0$ and ${\cal A} >
0$, and a change in sign of $\alpha$ signals the transition from
stability of the upright vertical stance $y = 0$ to instability or
falling : $y > 0$. This can be seen quite clearly by examining the
properties of the effective mechanical potential $U(y)$ associated
with our equation. This potential is obtained by integrating the
$y$-dependent force terms appearing in (\ref{nonlinear}) and yields
\begin{equation}\label{potential}
U(y) = \frac{1}{2}\alpha y^2 + \frac{1}{3} \epsilon y^3,
\end{equation} 
since $F(y) = -U'(y)$ is the deterministic (non-random) force acting
on the string.  A plot of $U$ versus $y$ is shown for $\alpha > 0$ and
for $\alpha < 0$ in Figure 5, where $U(y)$ is plotted along the
vertical and $y$ along the horizontal axes, respectively.
\begin{figure}[b]
\epsfysize=10cm 
{\centerline{\epsfbox{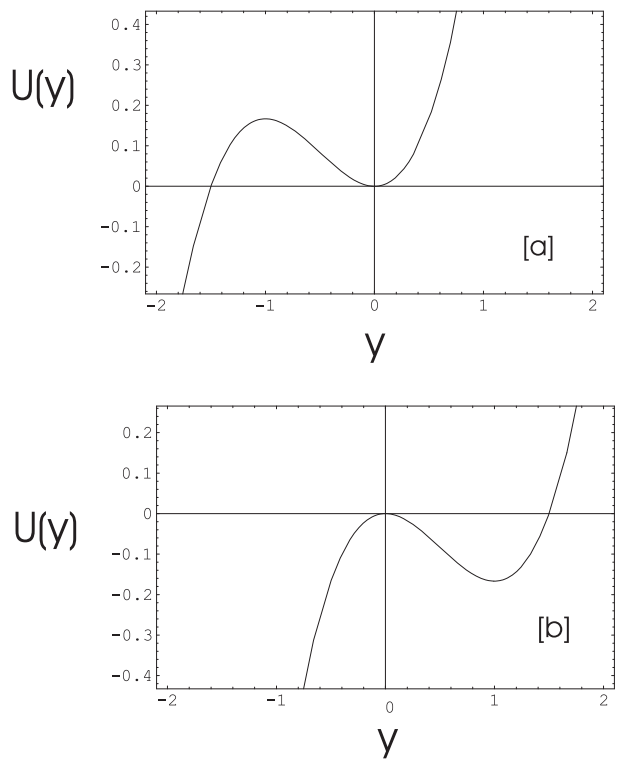}}}
\caption[]{\label{pin16.fig5}
Effective mechanical potential $U(y)$ for [a] $\alpha > 0$ 
and [b] $\alpha < 0$. $U(y)$ has units of energy and $y$ has units of length }
\end{figure}
For $\alpha > 0$, the origin at $y=0$ is locally stable to small
perturbations (to fall, one would have to be pushed backwards
sufficiently hard so as to overcome the potential barrier at $y = -1$.
However, when $\alpha < 0$, the vertical stance becomes an unstable
configuration, and a fall in either direction results (a clear
mechanical example of symmetry breaking).  This falling phenomena is
beyond the scope of the linear model since $\alpha$ is always positive
in the linear model.  The flow graphs in Section V reveal that $P2$
lies in a domain or phase which is separated from the domain or phase
of upright stance. Thus, healthy individuals are characterized by
having their initial conditions of posture control the basin of
attraction of $L1$.

The above group of dimensionless couplings forms a useful set when
$\nu$ is small, and the limit as $\nu$ tends to zero (corresponding to
vanishing viscosity or diffusion) can be safely studied. The limit of
small $\beta$ can also be treated with this set.  When $\alpha$ is
small however, corresponding to the diffusive regime, a different set
of couplings is required.  To study the flow corresponding to $\alpha
\rightarrow 0$ (corresponding to a vanishing linear restoring force)
we introduce the set of dimensionless variables
\begin{equation}\label{dimensionless2}
G = g/h^3 = \frac{\epsilon^2 {\cal A}}{\pi \nu^3 \Lambda^5} ,\qquad
H = h^{-1} = \frac{\alpha}{\nu \Lambda^2}, \qquad
F=f = \alpha \beta.
\end{equation}
In terms of these new variables, the original set of RG flow equations
(\ref{RGEs}) takes the following form:
\begin{eqnarray}\label{RGE2}
\frac{d\alpha}{d\ell} &=& \alpha \Big(z - \frac{2 G}{H(1+H)^2} \Big),\nonumber \\
\frac{d\beta}{d\ell} &=& \beta \Big(-z - \frac{G}{(1+H)^2}
\left\{ \frac{F}{H} + \frac{H}{2F(1+H)^2} \right\} \Big),
\nonumber \\
\frac{d{\cal A}}{d\ell} &=& {\cal A} \Big(z-2\chi -1 
+ \frac{G}{(1+H)^2}\left\{ \frac{F}{H} + \frac{1}{(1+H)} \right\} \Big),
\nonumber \\
\frac{d\nu}{d\ell} &=& \nu \Big( z - 2 + \frac{G}{(1+H)^4}\left\{
H-1 + \frac{F}{H}(3+2H-H^2) \right\}
 \Big),
\nonumber \\
\frac{d\epsilon}{d\ell} &=& \epsilon \Big(z + \chi +\frac{4 G}{(1+H)^3} \Big).
\end{eqnarray}
By setting $G = 0$, we once again recover the naive scaling laws given
by (\ref{naive}).  In terms of this group of three dimensionless
couplings, the RGE flow is characterized by
\begin{eqnarray}\label{RGE3}
\frac{dG}{d\ell} &=& G \Big(5 + \frac{G}{H(1+H)^4}[
4F(-2 - H + H^2) + 6H(2 + H)] \Big),\nonumber \\ 
\frac{dH}{d\ell} &=& 2H + \frac{G}{(1+H)^4}[-2-3H-3H^2+F(-3 -2H + H^2)], \nonumber \\
\frac{dF}{d\ell} &=& F \Big(-\frac{GH}{2F(1+H)^4}-\frac{2G}{H(1+H)^2}- 
\frac{FG}{H(1+H)^2}
\Big).
\end{eqnarray}
These result from differentiating (\ref{dimensionless2}) and using
(\ref{RGE2}).  Once again, the dependence on the two exponents
$z,\chi$ drop out of the equations for the dimensionless variables.
We solve for the complete set of real fixed points which we denote
collectively by $(G^*,H^*,F^*)$ and are displayed in the second column
of Table II.  We note that we reproduce two of the fixed points
obtained with the original variables, namely P2 and P5. Although their
coordinates as expressed with the variables $(G,H,F)$ are distinct
from those in the $(g,h,f)$ system, their exponents, eigenvalues and
eigenvectors are identical in both systems. Once again, we see that
$P2$ is an attractor for falling: in these variables $G^* > 0
\Longrightarrow \nu^* > 0$ so that $H^* < 0 \Longrightarrow \alpha^* <
0$.

We also discover a repulsive fixed line l1 located at $(0,0,F) =
(0,0,f)$. This fixed line was ``missed'' by the first group of
parameters.  The posture data in the diffusive phase selects a point
on this line, namely $(0,0,0.05)$.

We substitute these fixed points into the original set of RGE's
(\ref{RGE2}) to obtain the corresponding pair of critical exponents:
$z=z(G^*,H^*,F^*),\chi= \chi(G^*,H^*,F^*)$ at each fixed point:
\begin{equation}\label{exponents2}
z = \frac{2G^*}{H^*(1+H^*)^2} \qquad 
\chi = -\frac{2G^*(1 + 3H^*)}{H^*(1+H^*)^3}.
\end{equation}
These follow from taking the equations for $\epsilon$ and $\alpha$.
Taking other pairs of RGE's yields expressions for the exponents that
evaluate to the same numerical values at the fixed points. In analogy
with the first set of RG equations, we calculate the eigenvalues and
eigenvectors associated with the linearization of (\ref{RGE3}) about
each fixed point (fixed line) and determine the nature of each point
from the eigenvalues. This information is organized below in Table II.

\begin{table}
\begin{center}
\begin{tabular}{|c|c|c|c|l|l|} \hline
{\rm Fixed}      & {\rm Position}  &Critical exponents&{\rm IR-eigenvalues} &{\rm IR-eigenvectors}  & { \rm Class} \cr
{\rm Point}      &  $(G*,H*,F*)$   &$(z,\chi)$        &                     &                       &               \cr
\hline 
\hline 
l1               & $(0,0,F)$       &  $(2,1/2)$         &$ (5,2,0) $         &$V^{1}_{L1}= (1,0,0)$  &{\rm Unstable{ }line{}} \cr
                 &                 &                    &                    &$V^{2}_{L1}= (0,1,0)$  &                      \cr
\hline           
P2 &$(0.04, -0.46, -1.79)$   &$(-0.56, -0.41)$ &$(-6.79, -0.67 \pm 0.27i )$  &$ V^{1}_{P2}=(0.16,-0.98,-0.13)$& {\rm Stable spiral} \cr
                 &            &                 &                            &$ V^{2}_{P2}=(0.05 - 0.02i, 0.25-0.30i, 0.92 )$&     \cr
   &             &            &                                              &$ V^{3}_{P2}=(0.05 + 0.02i, 0.25 + 0.30i, 0.92 )$&     \cr 
   &             &            &                                              &                                                 &     \cr
P5 &$(-0.11,-0.34,-0.07)$&$(1.48, 0.03)$       &$(-8.24, 1.50, -1.28)$       &$ V^{1}_{P5}=(-0.72,-0.69,0.09)$&{\rm Saddle point}  \cr
    &            &                             &                             &$ V^{2}_{P5}=(-0.39,0.88,0.26)$ &                    \cr
    &            &            &                             & $ V^{3}_{P5}=(0.19,0.12,0.97) $&                   \cr
\hline
\end{tabular}
 \caption{}{\label{tab:two}Fixed Points in terms of ${G,F,H}$.}
\end{center}
\end{table}

%----------------------------------------------------------------------------
\section{Correlation function: Scaling properties}
%----------------------------------------------------------------------------

As discussed in the Introduction, the RG fixed point analysis can be
used for predicting the asymptotic, large-distance and long-time
limits of the two-point correlation function of transverse
displacements from the vertical upright position:
\begin{equation}\label{fullcorr}
C(x_1- x_2, t_1 - t_2) = \langle (y(x_1,t_1) - y(x_2,t_2))^2
\rangle,
\end{equation}
which measures the fluctuations in the difference of transverse
displacements at two different points along the body and/or at two
different times.  Because the model is translationally invariant, this
function depends only on the differences $x_1 - x_2$ and $t_1 - t_2$
in space and in time. In the scaling regime, which holds when the
system is in the vicinity of one of its fixed points, it is easy to
derive the exact scaling behavior of (\ref{fullcorr}) which emerges in
the large distance and long time limits. Put $x = x_1 - x_2$ and $\tau
= t_1 - t_2$. Consider correlations in the time-domain measured at the
same point on the body, so that $x = 0$. Then from $y(x,\tau) =
s^{-\chi} y(sx, s^z\tau)$ we have that
\begin{equation}\label{temporal}
C(0,\tau) = s^{-2\chi}C(0,s^z\tau) \sim \tau^{{2\chi}/{z}} \, C(0,1) = 
B \,\tau^{{2\chi}/{z}},
\end{equation}
with $B$ a constant, which follows from choosing $s$ such that $s^z
\tau = 1$.  Next, for correlations in the spatial domain, set $\tau =
0$ and we obtain
\begin{equation}\label{spatial}
C(x,0) = s^{-2\chi}C(sx, 0) \sim x^{2\chi} \, C(1,0) = 
A \, x^{2\chi},
\end{equation}
which follows from choosing $s$ such that $sx = 1$; $A$ is a constant.
These results are consistent with and imply the scaling limits of the
scaling function $\Psi$ appearing above in (\ref{corrscaling}).

Using the Tables \ref{tab:one}, \ref{tab:two}, it is a simple matter
to calculate the scaling of the correlation function about each fixed
point.  First, consider the limiting case represented by ``turning''
off the quadratic interaction term: $\epsilon \rightarrow 0$.  In this
limit, we recover the linear model of Chow and Collins
(\ref{ChowCollins}) and we reproduce the corresponding scaling
behavior(s) of the correlation function.  In this limit (and for
uncorrelated noise) there are then only the two trivial fixed points
(actually, fixed lines) $l1,L1$ (see Table \ref {tab:one}),
corresponding to the small $\alpha$ and small $\nu$ limits,
respectively.  Reading off the critical exponents from Table
\ref{tab:one} and using (\ref{temporal}) we confirm that in the
neighborhood of $l1$ the correlation function scales as $C(\tau) = B
\tau^{1/2}$. (In \cite{Chow-Collins} the temporal correlations are
parametrized as $C(\tau) \sim \tau^{2H}$, so this corresponds in our
notation to $H = 1/4$, since evidently, $2H=1/2 $.) This exponent,
$2H=1/2$, falls well within the experimental range of $0.52 \pm 0.12$,
which is exactly equivalent to the exponent $H=1/4$ falling within the
experimental range $0.26 \pm 0.06$ which is the one cited in
\cite{Chow-Collins}.  In the neighborhood of $L1$, we see that $z=0$,
indicating the relaxation time is independent of length scale and is
indicative of a saturated regime.  Care must be taken in calculating
the correlation function since the resulting exponent in
(\ref{temporal}) is formally divergent. If we write the correlation
function (\ref{fullcorr}) in terms of the auto-correlation function
$C(\tau) = 2[S(0) - S(\tau)]$ and use the scaling of the latter as
derived in (\ref{corrscaling}) then we see that $S(\tau) \sim \lim_{z
  \rightarrow 0} \tau^{-1/z} = 0$, so that $C(\tau) \rightarrow 2S(0)
\sim \tau^0 = {const.}$ This reproduces the scaling obtained by Chow
and Collins by other means.  Unfortunately, the error in the null
exponent for the saturated regime is unknown.  These two power laws
correspond to the two scaling regions termed as ``diffusive'' and
``saturated'', respectively. Note moreover that $l1$ is repulsive and
therefore unstable to the slightest perturbation (fluctuation) while
$L1$ is stable and attractive.  These lines lie within a common
two-dimensional RG flow domain (in the $h-f$-plane), and the
transition from diffusive to saturated behavior is understood from the
RG point-of-view as a cross-over phenomenon; see Fig. 6. This plot
shows $l1$ to be at a finite distance in $h$ above $L1$.  The RG flow
as determined from the linear model starts from a point on $l1$ and
drops down vertically until it ends up on $L1$. The locus of all
possible flows ``starting'' from points on $l1$ forms a ``curtain''
which is depicted in this Figure. We superpose the complicated RG flow
due to the non-linear interaction for sake of comparison.  Actually,
while $g^* = G^* = 0$ is consistent with zero coupling $\epsilon = 0$
and $f^* = F^*$ holds simultaneously, there is in fact an inverse (and
singular) relation between $h$ and $H$, namely, $H = 1/h$. Since $L1$
has $h^* = 0$, the other line, $l1$, when plotted in terms of these
coordinates $(g,h,f)$, is actually infinitely far away from $L1$ This
we interpret as an artifact of the model, which remember, is based on
an {\it infinitely} long string.  Thus, for an infinitely long body,
the transition to the saturated or bounded phase would never take
place (the crossover time would be infinite). But this makes perfect
physical sense since in fact saturation is a finite-size effect. This
connection between finite system size and saturation is also drawn for
phenomena in surface growth phenomena \cite{Barabasi}.  Real bodies of
course are finite in size, and these two fixed lines would be
separated by a finite distance in $h$.

\begin{figure}[b]
\epsfysize=5cm 
{\centerline{\epsfbox{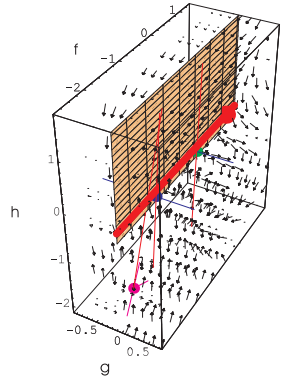}}}

\caption[]{\label{pin11}
Renormalization group flow between $l1$ and $L1$ in the $h-f$ plane.}
\end{figure}

This cross-over time scale is given by $\alpha^{-1}$, and a fit to the
data yields $\alpha^{-1} \approx 10 {\rm sec}$ \cite{Chow-Collins}.

Restoring the nonlinear term $\epsilon > 0$ gives rise to additional
nontrivial structure in RG parameter space.  In total we have the two
trivial fixed lines $(l1,L1)$ plus the four nontrivial fixed points:
$P2,P3,P4,P5$. The scaling behavior of the correlation function in
both the time and space domains are listed below in Table
\ref{tab:three}.

\begin{table}
\begin{center}
\begin{tabular}{|c|l|l|} \hline
Fixed Point& $C(\tau) = B\, \tau^{2\chi/z}$ &  $C(x) = A x^{2\chi}$ \cr 
\hline\hline 
l1 & $\tau^{\frac{1}{2}}$ & $x^1$ \cr
L1 & $\tau^0$ & $x^{-1}$ \cr
P2 & $\tau^{1.5}$ & $x^{-0.8}$ \cr
P3 & $\tau^{-5.9}$ & $x^{-0.8}$ \cr
P4 & $\tau^{-6.0}$ & $x^{-0.9}$ \cr
P5 & $\tau^{0.04}$ & $x^{0.1}$ \cr
\hline
\end{tabular}
\caption{}{\label{tab:three} Scaling behavior of the correlation function about each fixed point.}
\end{center}
\end{table}
The scaling of $C(\tau)$ associated with the trivial fixed lines $l1$
and $L1$ reproduces that obtained previously in \cite{Chow-Collins}
where the (bare) correlation function was calculated directly in the
linear model.  The behavior of $C(\tau)$ in the vicinity of the
non-trivial fixed points $P2,P3,P4$ and $P5$ is due entirely to the
presence of the non-linear term $\sim y^2$ in the equation of motion.
The scaling behavior in the {\it spatial} fluctuations encoded in
$C(x)$ was not discussed in \cite{Chow-Collins} but these can be
computed just as easily as the temporal fluctuations and are listed in
the third column of Table {\ref{tab:three}} for completeness. These
together with the behavior of $C(\tau)$ for the non-trivial fixed
points constitute predictions of the (non-linear) model within all the
phases.  We draw particular attention to the scaling of the
correlation function in the vicinity of the stable spiral $P2$, which
as we demonstrated in the previous Section, is an attractor for
falling.

%----------------------------------------------------------------------------
\section{Renormalization group Flow}
%----------------------------------------------------------------------------%

The two-dimensional RG flow for the strictly linear model, Fig 6., is
rather featureless and uniform and lends itself to easy
interpretation.  Although we have depicted it in the two-dimensional
coupling plane $h-f$, as already pointed out, the phenomenology of
posture control picks out a single point on each of the fixed lines
$l1$ and $L1$, $(0,0,F^*)$ and $(0,0,f^*)$, respectively, and the RG
flow is actually a one-dimensional line.  By marked contrast, the RG
flow for the nonlinear model fills out the full three-dimensional
coupling space $(g,h,f)$. The relative locations of $L1,P2,P3,P4,P5$
as shown in Fig. 7, where we have suppressed the flow for better
visibility. The largest point on the heavy line $L1$ represents the
experimentally determined point $P_{CC} = (0,0,0.05)$, and one can
appreciate the close proximity of the two saddle points $P3$ and $P4$,
in the neighborhood of this point. The stable spiral $P2$ is further
away and the other saddle point $P5$ is furthest removed from $L1$.
The actual numerical coordinates of all these objects are listed in
Table {\ref{tab:one}}. We have found the RG flow is best represented
in terms of the (normalized) instantaneous direction field, obtained
from directly plotting the differential equations
(\ref{dimensionless1}) as functions of $(g,h,f)$. In normalizing, we
lose information of the instantaneous ``speed'' of the flow, but
retain the sense of flow direction and flow morphology. The
instantaneous magnitudes of the flow vectors changes abruptly and
dramatically and without normalization, makes the graphs extremely
difficult to plot. We plot each fixed point with its local system of
numerically determined eigenvectors. The lengths and directions of
each eigenvector system is taken from Table \ref{tab:one}.  The
remaining graphs represent a selection of direction flow fields
calculated in the vicinity of the fixed points, for various ranges of
the coupling $g$, where the complicated nature of the RG flow can be
best appreciated.

\begin{figure}[b]
\epsfysize=5cm 
{\centerline{\epsfbox{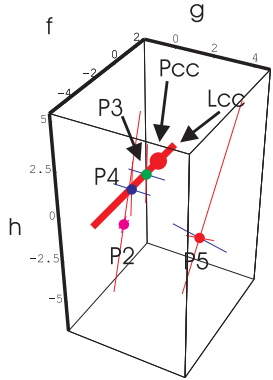}}}

\caption[]{\label{pin6}
The fixed line $L1$, experimental point $P_{CC}$, and  $P2,P4,P4,P5$.
The plotted eigenvectors are taken from Table \ref{tab:one}.}
\end{figure}

The structure of the RG flow is shown in Fig. 8, which gives a
``close-up'' view of the flow in the vicinity of the experimental
point and the saddle point $P4$. Gross feature can be appreciated such
as the circulation coming from the region of positive $f$ and sudden
change of flow in the region of negative $f$, which we interpret as
signaling the presence of a phase boundary or domain. In the closer
vicinity of the fixed points the flow is being attracted to the point
CC. Since $P4$ is a saddle point, the flow does not end up there, but
gets ``deflected'' when it passes by.

\begin{figure}[b]
\epsfysize=5cm 
{\centerline{\epsfbox{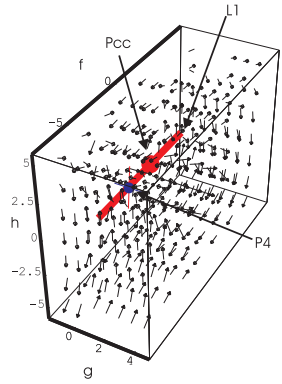}}}
\caption[]{\label{pin15}
The RG flow in the vicinity of CC and $P4$.}
\end{figure}

In Fig. 9, we show the structure of the RG flow in the vicinity of CC and 
the stable spiral $P2$.

\begin{figure}[b]
\epsfysize=5cm 
{\centerline{\epsfbox{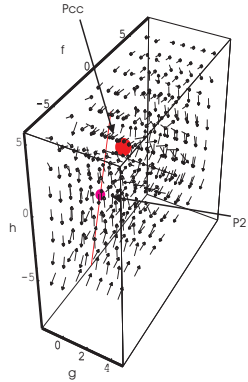}}}

\caption[]{\label{pin17}
The RG flow in the vicinity of CC and the stable spiral $P2$.}
\end{figure}

In Fig 10, we view the same flow field as before, but drawing the
points $P3$ and $P4$ in their respective positions, and Fig 11. gives
the view but with extended range in the coupling $g$, while in Fig 12,
the range in $g$ is extended even further and only the points $P_{CC}$
and $P3$ are drawn in place.

\begin{figure}[b]
\epsfysize=5cm 
{\centerline{\epsfbox{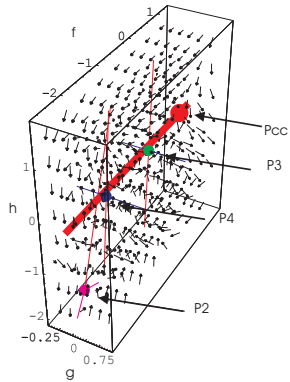}}}

\caption[]{\label{pin9}
The RG flow in the neighborhood of CC, $P3$,$P4$ and $P2$.}
\end{figure}

\begin{figure}[b]
\epsfysize=5cm 
{\centerline{\epsfbox{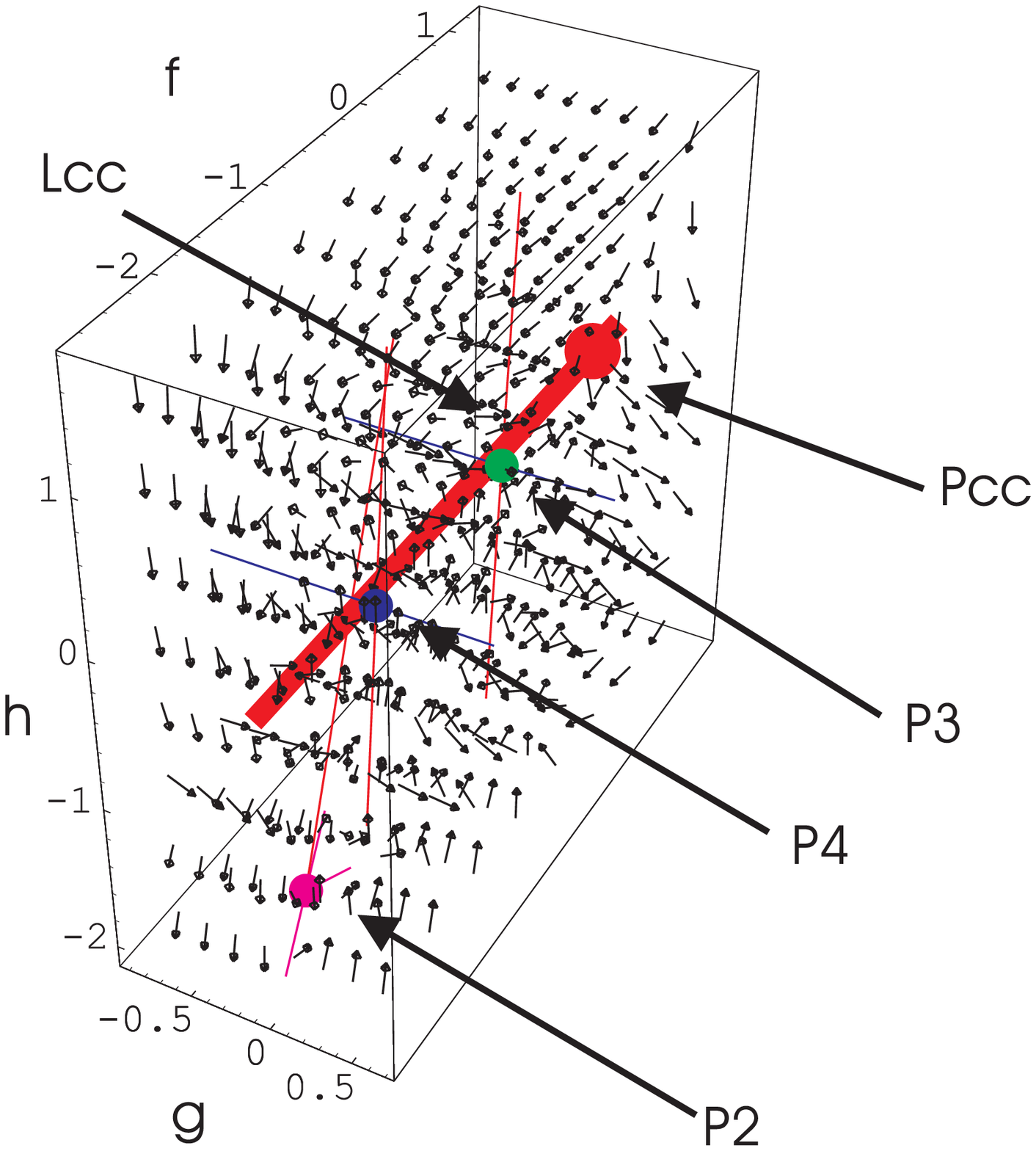}}}
\caption[]{\label{pin2}
The RG flow in the neighborhood of CC, $P3$, $P4$, and $P2$. }
\end{figure}

\begin{figure}[b]
\epsfysize=5cm 
{\centerline{\epsfbox{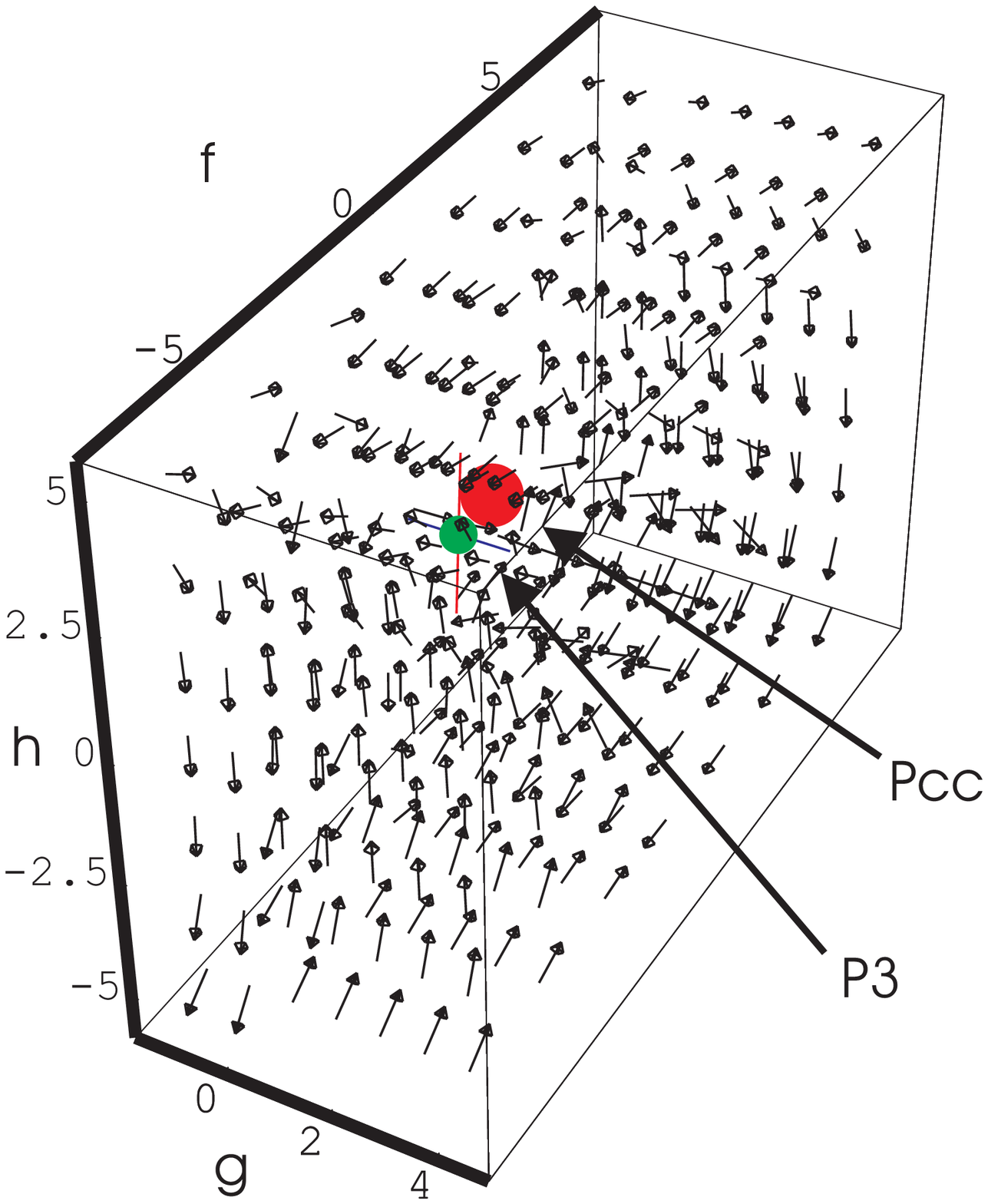}}}
\caption[]{\label{pin18}
The RG flow in the neighborhood of CC and $P3$.}
\end{figure}

\begin{figure}[b]
\epsfysize=5cm 
{\centerline{\epsfbox{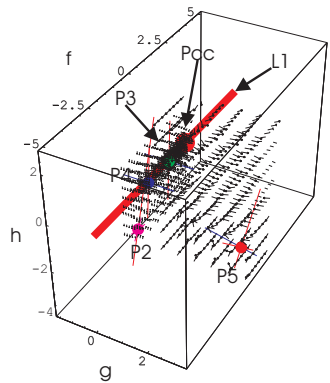}}}
\caption[]{\label{pin11}
All points and flow field are included.}
\end{figure}

%-------------------------------------------------------------------
\section{Summary and Discussion}
%-------------------------------------------------------------------

We have extended Chow and Collin's linear pinned-polymer model of
posture control by including a weak quadratic non-linearity. There are
at least two good reasons for doing so. Firstly, in real
anteroposterior movement, the front to back sway is not symmetric with
respect to the vertical upright position.  An obvious way to account
for this fact is to introduce a symmetry-breaking term in the polymer
equation of motion, and the minimal term that can be added is second
order or quadratic in the displacement field $y$.  Secondly, the
effects such as stepping or falling are beyond the scope of the linear
model, and these can be approximately modelled by means of non-linear
terms in the equation of motion \cite{Koleva}.  The analysis of
non-linear equations is a complicated enterprise, but the techniques
afforded by the renormalization group permit one to obtain a wealth of
information regarding the dynamical phases of the system for both
large-distance and long-time limits. We have undertaken a detailed RG
analysis of the fully nonlinear model and have summarized our results
in terms of RG fixed points, stability analysis and exponents.  We
have also numerically computed the fully non-linear RG flow and have
represented this in terms of a normalized vector flow-field.
Knowledge of the RG fixed points is sufficient for determining the
exact power-law behavior of the correlation function of posture
displacement in both the temporal and spatial domains.  In the linear
limit of the model we recover the diffusive and saturated phases of
posture control and compute the scaling of the correlation function
within each phase.  The transition from the diffusive to the saturated
phase and the associated change in scaling exponent is a crossover
phenomena.  The crossover time is however, finite, only for finite
size systems, as we have argued.  These results agree with the linear
analysis of Chow and Collins \cite{Chow-Collins}.  The quadratic
non-linearity gives rise to four non-trivial fixed points. There are
two saddle points near the attractive trivial fixed point. While they
do alter the RG flow in the neighborhood of the trivial attractive
fixed point, they have no bearing on either of the two linear phases
(diffusive and saturated) of the model. The linear model fits the
posture data rather well, and the detailed analysis undertaken here
substantiates the claim made in \cite{Chow-Collins} that weak
nonlinearities are not needed to explain the posture data of healthy
individuals.

The other two non-trivial fixed points consist of an additional saddle
point and a stable spiral. The saddle point corresponds to a large
value of the dimensionless coupling constant and is probably an
artifact of perturbation theory. Much more interesting is the spiral
which is purely attractive and as we have shown, is associated with a
falling phase. Its domain of attraction appears to be separated from
the diffusive and saturated domains. It is important to note that the
quadratic nonlinearity is the minimum term that can be added to the
equation of motion that serves to {\it break} the anteroposterior
symmetry. This symmetry breaking has lead to a falling phase, and thus
this one term simultaneously fulfills two distinct requirements. In
\cite{Koleva}, falling has been modelled with a nonlinear potential,
but the nonlinearity employed there does not break the $y \rightarrow
-y$ symmetry.

Mention should be made of the short-time inertial effects which are
not covered in the present analysis, and which are important for an
understanding of posture control. These seem to imply the existence of
short-time correlations in the noise \cite{Chow-Collins} and is no
surprise therefore that the purely white uncorrelated noise used here
is unable to reproduce this early scaling regime. Nevertheless, the
analysis carried out here can be straightforwardly extended to handle
both white and colored noise, and some comments to this effect are
provided in Appendix D.

Although our analysis has centered on the application to posture
control, variants of stochastic differential equations of the type
considered here have applications to a host of other problems where
nonlinear waves propagate in a noisy and/or random medium
\cite{KonotopVazquez}, and the general details of the RG analysis
carried out here should be useful for addressing these other
applications.

%-------------------------------------------------------------------
\section*{Acknowledgments}
%-------------------------------------------------------------------
We thank David R.C. Dom\'\i nguez for many interesting discussions
during the early stages of this work and for supplying us with
independent numerical integrations of the differential RG equations.
This work was financed in part by INTA, FSE (Fondo Social Europeo),
FEDER, CSIC (Consejo Superior de Investigaciones Superiores), CAM
(Comunidad Aut\'onoma de Madrid) and OCYT (Oficina de Ciencia y
Tecnolog\'\i a de la Presidencia del Gobierno).

%---------------------------------------------------------------------------
\appendix
%---------------------------------------------------------------------------
\section{Response function: one-loop correction}
%---------------------------------------------------------------------------

The explicit analytic expression for the one-loop correction
(hereafter denoted by primes) to the response function
(\ref{response}) follows immediately from transcribing the
diagrammatic representation of the corrected response function (see
Fig. 2) into its corresponding mathematical elements:
\begin{eqnarray}\label{oneloopr}
\Delta'(p,\omega) &=& \Delta(p,\omega) + 4\epsilon^2 \Delta(p,\omega)\times
I_r(p,\omega)\times \Delta(p,\omega) \nonumber \\
&=& \Delta(p,\omega)\big(1 +  4\epsilon^2 \Delta(p,\omega)\times
I_r(p,\omega) \big)\nonumber \\
\Rightarrow \Delta'^{-1}(p,\omega) &=& \Delta^{-1}(p,\omega) - 4\epsilon^2 
I_r(p,\omega) + O(\epsilon^4),
\end{eqnarray}
where the loop-integral $I_r(p,\omega)$ is built up from the bare
response function, the bare vertex and the bare noise spectrum (for
convenience, we have already factored the dependence on the vertex, or
bare coupling $\epsilon$, out of the loop integral).  {F}rom
inspection of the loop diagram and making use of the Feynman rules,
this integral has the structure given by
\begin{eqnarray}\label{responseloop}
I_r(p,\omega) &=& \int^{>} \frac{dq}{2\pi}\int_{-\infty}^{\infty}\,
\frac{d{\Omega}}{2\pi}\, \Delta(p-q,\omega-\Omega)\,C(q,\Omega),\nonumber \\
&=&\int^{>} \frac{dq}{2\pi}\int_{-\infty}^{\infty} 
\frac{d{\Omega}}{2\pi}\,
\frac{\Gamma(q,\Omega)}{\big(\Omega^2 + [\nu q^2 +\alpha -\beta \Omega^2]^2\big)
\big(i[\omega-\Omega]
-\beta[\omega-\Omega]^2 + \nu(p-q)^2 + \alpha \big)},
\end{eqnarray}
valid for an arbitrary Gaussian noise spectral function $\Gamma(q,\Omega)$. 

The internal momentum and frequency flowing around the loop are
denoted by $q$ and $\Omega$, respectively. The net momentum and
frequency flowing into and out of the loop-diagram is $p$ and
$\omega$; note that conservation of momentum and frequency is
maintained independently at each vertex.  We take a white noise
spectrum $\Gamma(q,\Omega) = 2{\cal A}$, and first compute the
frequency integral exactly using the residue theorem (the contour may
be closed in either the upper or lower half-plane).  This yields for
the one-loop corrected inverse response function (from the last line
in (\ref{oneloopr}))
\begin{equation}\label{responseprimed}
\big(i\omega -\beta' \omega^2 + \nu' p^2 + \alpha'\big) =
\big(i\omega -\beta \omega^2 + \nu p^2 + \alpha \big) + 
\frac{4\epsilon^2 {\cal A}}{\beta^2}
\int^{>} \frac{dq}{2\pi}\, F(q;p,\omega),
\end{equation}
where the integrand function $F$, which depends on both internal and external
momenta as well as on the external frequency, is given by 
\begin{equation}\label{F}
F(q;p,\omega) = \frac{1}{(\Omega_1-\Omega_2)}
\left\{ \frac{1}{\Omega_1(\Omega_1-\Omega_3)(\Omega_1-\Omega_4)} +
\frac{1}{\Omega_1^*(\Omega_2-\Omega_3)(\Omega_2-\Omega_4)} \right\},
\end{equation}
and is expressed in terms of the following poles
in the complex frequency plane which arise in the frequency integration: 
\begin{eqnarray}\label{poles}
\Omega_1 &=& \frac{1}{2\beta}\Big(i + \sqrt{4\beta(\nu q^2 + \alpha) - 1} \Big),\nonumber \\
\Omega_2 &=& \frac{1}{2\beta}\Big(i - \sqrt{4\beta(\nu q^2 + \alpha) - 1} \Big),\nonumber \\
\Omega_3 &=& \frac{1}{2\beta}\Big(-i + \sqrt{4\beta[\nu (p-q)^2 + \alpha] - 1} \Big) + \omega, 
\nonumber \\
\Omega_4 &=& \frac{1}{2\beta}\Big(-i - \sqrt{4\beta[\nu (p-q)^2 + \alpha] - 1} \Big) + \omega. 
\end{eqnarray}

We work with the {\it inverse} response function since this is a
simple polynomial in $p$ and $\omega$.  To renormalize $\Delta^{-1}$,
we must expand out the momentum-integral in (\ref{responseprimed}) in
lowest powers in both the external frequency and momentum
$(\omega,p)$, match like-powers on both sides of the expression
(\ref{responseprimed}) and then take the hydrodynamic limit $\omega
\rightarrow 0$, $p \rightarrow 0$ at the end of the calculation.  It
is important to note that contributions to this asymptotic
long-distance and long-time expansion come not only from
Taylor-expanding the integrand $F$ itself but also from the {\it
  domain of integration} implicit in the integral.  We must integrate
the loop momentum within a fixed ``shell'' and the net momenta
circulating within the loop depends on both external and internal
momentum variables, and this fact must be taken into account. Thus,
the resultant domain of momentum-shell integration is given by the
{\it intersection} of the two intervals $\Lambda/s \leq |q| \leq
\Lambda$ and $\Lambda/s \leq |p-q| \leq \Lambda$. Up to second order
in $p$, the last inequality can be written as $\Lambda/s +
p\frac{|q|}{q} \leq |q| \leq \Lambda + p\frac{|q|}{q}$, since the
$O(p^2)$ terms vanish identically. In taking the intersection of this
with the first inequality, we have four cases to consider depending on
the sign of $p$ ($p > 0, p < 0$) and the sign of $\frac{|q|}{q} =
\pm$. The resultant integration domain, valid for all four cases, can
be written as the difference
\begin{equation}\label{domain}
\int^{>} \frac{dq}{2\pi} = \int_{\Lambda/s}^{\Lambda} \frac{dq}{2\pi} -
\int_{\Sigma(p,s)} \frac{dq}{2\pi},
\end{equation} 
where the domain $\Sigma(p,s) = [\Lambda/s,\Lambda/s + p] \cup [\Lambda + p,\Lambda]
\cup [\Lambda - p,\Lambda] \cup [\Lambda/s, \Lambda/s - p]$. Note of course that
$\Sigma(0,s) = \Sigma(p,1) = \phi$, is just the empty set.

To proceed with the calculation, in accord with
(\ref{responseprimed}), we need to expand out the function $F$ up to
and including quadratic powers in both external frequency and momentum
$(1, \omega, p, p^2, \omega p, \omega^2)$ and consistently combine
these with the powers of $p$ coming from the integration over
$\Sigma(p,s)$.  In practice, this delicate operation need only be
carried out for the renormalization of the diffusion constant $\nu$.
This is because the parameter $\alpha$ does not multiply any positive
power of either frequency or momentum, so we can take the hydrodynamic
limit at the outset in computing its one-loop correction. Next, the
parameter $\beta$ multiplies $\omega^2$, so we must expand the
integrand $F$ to this same order to obtain the correction $\beta'$,
but we can set the external momentum $p$ to zero at the outset: the
domain $\Sigma(p,s)$ does not depend on external frequency and makes
no contribution to the renormalization of $\beta$.  Terms linear in
external frequency $(\omega)$ appearing in the loop integral do not
yield any new information since we can always re-define the time to
absorb such corrections when they arise (thus we maintain the unit
coefficient $1$ in front of the term $i\omega$ in $\Delta'$).
Finally, for the viscosity renormalization, we can set the external
frequency to zero at the outset but must expand the integrand together
with the integration domain up to and including second order in the
external momentum $(p)$. The constant contribution serves to
renormalize $\alpha$, as we have already remarked. Taking these points
into consideration, we arrive at the following one-loop expressions
for $\alpha',\beta',\nu'$:
\begin{eqnarray}
\alpha' &=& \alpha + 4\frac{\epsilon^2 {\cal A}}{\beta^2} \int_{\Lambda/s}^{\Lambda}
\frac{dq}{2\pi} \, F(q,0,0),\nonumber \\
\beta' &=& \beta -4\frac{\epsilon^2 {\cal A}}{\beta^2} \int_{\Lambda/s}^{\Lambda}
\frac{dq}{2\pi} \frac{1}{2!}\, \frac{\partial^2 F(q,0,0)}{\partial \omega^2},\nonumber \\
\nu'\, p^2 &=& \nu \, p^2 + 4\frac{\epsilon^2 {\cal A}}{\beta^2}
\Big(p^2 \int_{\Lambda/s}^{\Lambda} \frac{dq}{2\pi} \,
\frac{1}{2!}\frac{\partial^2 F(q,0,0)}{\partial p^2}
-p \int_{\Sigma(p,s)} \frac{dq}{2\pi} \, \frac{\partial F(q,0,0)}{\partial p} \Big).
\end{eqnarray}

Calculating the indicated derivatives of $F$ using the definition of
(\ref{F}) together with the complex poles (\ref{poles}) yields the
following independent equations for the (one-loop) renormalized
parameters, namely
\begin{eqnarray}\label{responseprime}
\alpha' &=& \alpha - 2\epsilon^2 {\cal A} \int^{\Lambda}_{\Lambda/s} \frac{dq}{2\pi}\,
\frac{1}{(\nu q^2 + \alpha)^2},\nonumber \\
\beta' &=& \beta - \epsilon^2 {\cal A} \int^{\Lambda}_{\Lambda/s} \frac{dq}{2\pi}\,
\Big(\frac{\beta^2}{(\nu q^2 + \alpha)^2} + \1/2
\frac{1}{(\nu q^2 + \alpha)^4} \Big),\nonumber \\
\nu' &=& \nu + 2 \epsilon^2 {\cal A} \int^{\Lambda}_{\Lambda/s}
\frac{dq}{2\pi}\,
\frac{1}{[4\beta(\nu q^2 + \alpha) - 1]} \left\{
[\frac{2\nu}{\beta}-\frac{8\nu^2 q^2}{4\beta(\nu q^2 + \alpha) - 1}]
\Big(-\frac{\beta^3}{(\nu q^2 + \alpha)} + \frac{5\beta^2}{4(\nu q^2 + \alpha)^2}
-\frac{\beta}{4(\nu q^2 + \alpha)^3} \Big) \right. \nonumber \\
&+&  \left. 4\nu^2 q^2 \Big(\frac{3\beta^2}{2(\nu q^2 + \alpha)^2}
-\frac{\beta}{(\nu q^2 + \alpha)^3} + \frac{1}{4(\nu q^2 + \alpha)^4}\Big) \right\}
+ O(\Sigma(p,s)).
\end{eqnarray}

These one-loop equations are exact. We have not bothered to explicitly
write out the contribution to the viscosity renormalization coming
from the domain $\Sigma(p,s)$, since it is easy to show that this will
vanish identically when we pass to the differential form of the
renormalization group equations, i.e, in the limit of a thin-shell.
That is, for any function $f$, we have
\begin{equation}
\frac{d}{ds}|_{s=1} \int_{\Sigma(p,s)} f(u)\, du = 0.
\end{equation}

Applying the renormalization group procedure as described in Section
III (a Kadanoff transformation or coarse-graining followed by a
re-scaling) to these equations yields the corresponding differential
RG equations in (\ref{RGEs}).

%---------------------------------------------------------------------------
\section{Noise spectral function: one-loop correction}
%---------------------------------------------------------------------------

{F}rom the diagrammatic one-loop expansion for the correlation function 
we obtain the one-loop 1PI diagram representing the noise spectral
function corrected to one-loop, as shown in the diagram in Fig. 4. 
This translates into the following mathematical equation (after factoring out
the dependence on the bare vertex or coupling) for a general Gaussian noise
spectral function
\begin{equation}
\Gamma'(p,\omega) = \Gamma(p,\omega) + 2\epsilon^2\, I_n(p,\omega),
\end{equation}
which for the case of white noise considered here, reduces to 
\begin{equation}
2{\cal A}' = 2{\cal A} + 2\epsilon^2 \, I_n(p,\omega).
\end{equation}
The loop integral $I_n$ depends in general on external frequency and
momentum, whose structure is given by 
\begin{equation}
I_n(p,\omega) =  \int^{>} \frac{dq}{2\pi}\int_{-\infty}^{\infty}
\frac{d\Omega}{2\pi} \,\, C(q,\omega) C(p-q, \Omega-\omega),
\end{equation}
but since the  noise spectrum is constant, we can
take the hydrodynamic limit right away and evaluate
the somewhat simpler integral
\begin{equation}
I_n(0,0) = \int_{\Lambda/s}^{\Lambda} \frac{dq}{2\pi}
\int_{-\infty}^{\infty} \frac{d\Omega}{2\pi} \,
\frac{4 {\cal A}^2}{[\Omega^2 + (\nu q^2 - \beta \Omega^2 + \alpha)^2]^2}.
\end{equation}
Once again, the frequency integration can be evaluated easily by the method of residues.
Doing so, we obtain the one-loop correction to the white noise amplitude
\begin{equation}\label{noiseprime}
{\cal A}' = {\cal A} + 4 \epsilon^2 {\cal A}^2
\int_{\Lambda/s}^{\Lambda} \frac{dq}{2\pi} \,
\frac{1}{4\beta (\nu q^2 + \alpha)- 1}\Big(\frac{\beta^2}{(\nu q^2 + \alpha)}
+\frac{3}{4}\frac{\beta}{(\nu q^2 + \alpha)^2}-
\frac{1}{4(\nu q^2 + \alpha)^3} \Big).
\end{equation}

Applying the renormalization group procedure to this equation
yields the corresponding RG equation for the noise
amplitude given in (\ref{RGEs}). 

%----------------------------------------------------------------------------
\section{Vertex function: one-loop correction}
%----------------------------------------------------------------------------

The diagrammatic expansion for the one-loop correction to the vertex
function, or coupling constant, is depicted as shown in Fig. 3.  For
general vertex functions, momentum and frequency (i.e, energy)
conservation implies that a trilinear vertex can depend on at most two
independent external momenta and two independent external frequencies.
Which two momenta and which two frequencies one chooses is immaterial.
Translating the vertex diagrams into corresponding mathematical
elements yields for the one-loop vertex correction the following
equation:
\begin{equation}
\epsilon' = \epsilon + 4 \epsilon^3 I_v(k_1,\omega_1;k_2,\omega_2).
\end{equation}
The structure of the one-loop integral is given as follows:
\begin{eqnarray}
 I_v(k_1,\omega_1;k_2,\omega_2) &=& \int^{>} \frac{dq}{2\pi} \int_{-\infty}^{\infty}
\frac{d\Omega}{2\pi} \,\, \left\{C(q,\Omega)\Delta(q-k_2,\Omega-\omega_2)
\Delta(q-k_1,\Omega-\omega_1)\right.
\nonumber \\
&+& \Delta(-q,-\Omega)C(q-k_2,\Omega-\omega_2)\Delta(q-k_1,\Omega-\omega_1)\nonumber \\
&+&  \left. \Delta(-q,-\Omega)\Delta(k_2- q,\omega_2 - \Omega)
C(k_1 -q,\omega_1 - \Omega)\right\}.
\end{eqnarray}
However, since the coupling $\epsilon$ is constant, in anticipation of
the hydrodynamic limit we can immediately set all external momenta and
frequencies to zero in computing the one-loop correction to
$\epsilon$.  Taking this limit, and taking a white noise spectral
function, we have that the vertex loop integral at zero external
momentum and frequency is given by
\begin{eqnarray}\label{vertloop}
I_v(0,0;0,0) = \int^{>} \frac{dq}{2\pi} \, \int_{-\infty}^{\infty} \frac{d\Omega}{2\pi}\,
\frac{2{\cal A}}{ [\Omega^2 + (\nu q^2 -\beta \Omega^2 + \alpha)^2] } &\Big(&
\frac{1}{(i\Omega - \beta \Omega^2 + \nu q^2 + \alpha)^2} \nonumber \\
&+& \frac{1}{(i\Omega - \beta \Omega^2 + \nu q^2 + \alpha)
(-i\Omega - \beta \Omega^2 + \nu q^2 + \alpha)} \nonumber \\
&+& \frac{1}{(-i\Omega - \beta \Omega^2 + \nu q^2 + \alpha)^2} \Big).
\end{eqnarray}
The integral over the internal frequency may be performed exactly, once again by the
method of residues. This yields 
\begin{equation}\label{vertexprime}
\epsilon' = \epsilon + 8 \epsilon^3 {\cal A}
\int^{\Lambda}_{\Lambda/s} \frac{dq}{2\pi} \Big( \frac{1}{4(\nu q^2 + \alpha)^3}
-\frac{\beta}{4(\nu q^2 + \alpha)^2} +
\frac{1}{(4\beta(\nu q^2 + \alpha)-1)}\left\{
\frac{\beta^2}{(\nu q^2 + \alpha)}+\frac{3\beta}{4(\nu q^2 + \alpha)^2}
-\frac{1}{4(\nu q^2 + \alpha)^3} \right\} \Big).
\end{equation}

Applying the renormalization group procedure to this equation yields
the differential RG equation for the coupling listed in (\ref{RGEs}).

%----------------------------------------------------------------------------
\section{Correlated noise}
%----------------------------------------------------------------------------%

The effect of both temporal and/or spatial correlations in the
Gaussian noise spectrum can be also be taken into account in this
model. Here, we briefly indicate what steps would have to be taken or
modified in the renormalization group program to include such
correlations. In (\ref{noise}) the spectral function with both
uncorrelated (white) and correlated (colored) components is written
\begin{equation}
\Gamma(k,\omega) = 2{\cal A} + 2{\cal A}_{\rho,\theta}\big(\frac{k^2}{\Lambda^2}\big)^{-\rho}
\,\big(\frac{\omega^2}{\omega_0^2}\big)^{-\theta},
\end{equation}
where we consider long-range correlations of the power-law type. These
are parametrized in terms of two exponents, $\rho$ and $\theta$, for
spatially and temporally correlated noise, respectively. The naive
scaling properties (\ref{naive}) of the stochastic equation are
extended to include the scaling of the correlated part of the noise,
which reads
\begin{equation}
{\cal A}_{\rho,\theta} \rightarrow s^{2\rho - 1 -2\chi + z(2\theta + 1)}
{\cal A}_{\rho,\theta}.
\end{equation}

The perturbative expansion for the response, noise and vertex goes
through as before, except now, for $\theta \neq 0$, the one-loop
frequency integrations over $\Omega$ must be re-calculated and in
general, branch cuts and poles must be dealt with in the complex
$\Omega$-plane. The RGE's will now depend on the two noise exponents
$\rho$ and $\theta$, as will the fixed points and the critical
exponents: $z = z(\rho,\theta)$, and $\chi = \chi(\rho,\theta)$.
There will be an additional RGE for the amplitude of the colored
component of the noise yielding a total of six equations. By
Buckingham's $\Pi$-Theorem \cite{Buckingham}, we know that these can
be cast in terms of four equations in four dimensionless variables. In
effect, the correlations in the noise ``open up'' a new direction in
parameter space and yield a correspondingly more complicated fixed
point and RG flow structure than that of the uncorrelated noise case
treated here.

%-------------------------------------------------------------------

%-------------------------------------------------------------------

%-------------------------------------------------------------------
\end{document}